%% file: HNSecW-v2.tex
\documentclass[sigconf, nonacm, pdfa, balance=false]{acmart}
\usepackage{pvldb}
\usepackage[a-2b]{pdfx}
\makeatletter
\let\hyxmp@embed@packet\relax
\let\hyxmp@check@standards\relax
\let\hyxmp@warn@if@no@metadata\relax
\makeatother
\input{graphics}
\input{font-maps}
\usepackage{needspace}
\usepackage{flushend}

\renewcommand\vldbdoi{10.14778/3849423.3849440}
\renewcommand\vldbpages{284 - 297}
\renewcommand\vldbvolume{20}
\renewcommand\vldbissue{2}
\renewcommand\vldbyear{2026}
\renewcommand\vldbauthors{\authors}
\renewcommand\vldbtitle{\shorttitle}
\renewcommand\vldbavailabilityurl{https://github.com/Kleenex-ultrasoft/Secure_ANN}
\renewcommand\vldbpagestyle{empty}

\usepackage[color,secparam=$\kappa$]{crypto}
\usepackage{enumitem}
\usepackage{tabularx}
\usepackage{booktabs}
\usepackage{algorithm,xspace}
\usepackage{multirow}
\usepackage[noend]{algpseudocode} %
\usepackage{makecell}
\usepackage{amsmath}
\usepackage[normalem]{ulem} %
\usepackage{xcolor}
\algrenewcommand\algorithmicrequire{\textbf{Input:}}
\algrenewcommand\algorithmicensure{\textbf{Output:}}

\newcommand{\share}[1]{\ensuremath{\llbracket#1\rrbracket}}
\AtBeginDocument{%
  }

\usepackage{xspace} %

\providecommand{\Querier}{\ensuremath{\mathsf{Querier}}\xspace}

\usepackage{subcaption}

\setcopyright{none}
\acmYear{}  \copyrightyear{}
\acmISBN{}  \acmDOI{}
\makeatletter
\renewcommand\footnotetextcopyrightpermission[1]{}
\makeatother

\newcommand{\sysname}{{\textsf{HNSecW}}\xspace}

\begin{document}

\title{HNSecW: A Logarithmic-Time MPC Protocol for Approximate Nearest Neighbor Search}

\author{Fahong Zhang}
\affiliation{%
  \institution{Nanyang Technological University}
  \city{Singapore}
  \country{Singapore}}
\email{fahong001@e.ntu.edu.sg}

\author{Weijie Jin}
\authornote{The author took part in and made all contributions to this work during his visiting internship at NTU.}
\affiliation{%
  \institution{Nanyang Technological University}
  \city{Singapore}
  \country{Singapore}}
\email{WJIN003@e.ntu.edu.sg}

\author{Lixu Wang}
\affiliation{%
  \institution{Nanyang Technological University}
  \city{Singapore}
  \country{Singapore}}
\email{lixu.wang@ntu.edu.sg}

\author{Ning Luo}
\affiliation{%
  \institution{University of Illinois Urbana-Champaign}
  \city{Urbana}
  \country{USA}}
\email{nl27@illinois.edu}

\author{Xiao Wang}
\affiliation{%
  \institution{Northwestern University}
  \city{Evanston}
  \country{USA}}
\email{wangxiao@northwestern.edu}

\author{Wei Dong}
\authornote{Corresponding author.}
\affiliation{%
  \institution{Nanyang Technological University}
  \city{Singapore}
  \country{Singapore}}
\email{wei\_dong@ntu.edu.sg} %

\renewcommand{\shortauthors}{Zhang et al.}

\begin{abstract}
As a core component of modern RAG systems, approximate nearest neighbor (ANN) search has been extensively studied for decades. In practice, deploying ANN often requires symmetric privacy, where the querier’s query and the database owner’s data must remain hidden from each other, necessitating a secure multi-party computation (MPC) solution. So far, most MPC-based ANN protocols follow a cluster-based approach: they evaluate distances for the query over a large candidate set and rely on heavyweight primitives such as distributed ORAM to hide access patterns. As a result, each search typically requires $O(\sqrt{N})$ time. We propose \sysname, a novel MPC protocol for ANN search that achieves the $O(\log N)$ search time for the first time. We first study the \emph{single-query} setting. Starting from HNSW, the state-of-the-art ANN algorithm, we redesign its control flow and data structures to achieve fully data-independent execution. To further achieve logarithmic search time, we employ offline reshuffling and partial database duplication. Furthermore, we extend our protocol to support \emph{multiple sequential queries} by introducing a ``double visited'' caching structure that eliminates redundant cross-query memory accesses, thereby avoiding per-query online shuffling. In addition, we develop a \emph{batch-query} protocol that exploits higher parallelism to reduce both round complexity and the circuit’s critical path. Our experiments show that \sysname preserves the high accuracy of plaintext HNSW ($\geq 90\%$ recall) and delivers over a $20\times$ efficiency improvement over prior work on million-scale datasets.
\end{abstract}

\makeatletter
\let\savedauthorfnsymbol\@fnsymbol
\renewcommand{\@fnsymbol}[1]{%
  \ifnum#1=1\relax
    \savedauthorfnsymbol{2}%
  \else\ifnum#1=2\relax
    \savedauthorfnsymbol{1}%
  \else
    \savedauthorfnsymbol{#1}%
  \fi\fi}
\makeatother
\maketitle
\makeatletter
\let\@fnsymbol\savedauthorfnsymbol
\makeatother

\begingroup
\let\savedvspace\vspace
\renewcommand{\vspace}[1]{\savedvspace{#1}\Needspace{5\baselineskip}}
\vldbtopmatter
\endgroup
\pagestyle{empty}

\input{Chapters/1_Introduction.tex}

\input{Chapters/2_Preliminaries.tex}

\input{Chapters/3_System_Overview}

\input{Chapters/4_Single_Query}

\input{Chapters/5_Multiple_Query}

\input{Chapters/6_Batch_Query}

\input{Chapters/7_Evaluation}

\input{Chapters/8_Related_Work}

\input{Chapters/Acknowledgments}

\clearpage
\bibliographystyle{ACM-Reference-Format}
\bibliography{references}

\end{document}

%% file: graphics.tex
\hypersetup{hidelinks}
\newsavebox{\paperfigurebox}
\newcommand{\papergraphics}[2][]{%
  \sbox{\paperfigurebox}{\includegraphics[#1]{#2}}%
  \leavevmode%
  \usebox{\paperfigurebox}}

%% file: font-maps.tex
\pdfmapline{=LinBiolinumTI-lf-ly1--base HNSecWBiolinumTI " AutoEnc_3gqkkcsy65zvxq7sordtjoyxsd ReEncodeFont " <[lbtn_3gqkkc.enc <fonts/HNSecWBiolinumTI.pfb}
\pdfmapline{=LinBiolinumTI-lf-ot1--base HNSecWBiolinumTI " AutoEnc_zlmmkbkd32ihxyyk23ea4f3deb ReEncodeFont " <[lbtn_zlmmkb.enc <fonts/HNSecWBiolinumTI.pfb}
\pdfmapline{=LinBiolinumTI-lf-sc-ly1--base HNSecWBiolinumTI " AutoEnc_l4ygyhtoeji3qsxowkyjr5f32b ReEncodeFont " <[lbtn_l4ygyh.enc <fonts/HNSecWBiolinumTI.pfb}
\pdfmapline{=LinBiolinumTI-lf-sc-ot1--base HNSecWBiolinumTI " AutoEnc_avrf3yr76bdldbd7omptgnriwb ReEncodeFont " <[lbtn_avrf3y.enc <fonts/HNSecWBiolinumTI.pfb}
\pdfmapline{=LinBiolinumTI-lf-sc-t1--base HNSecWBiolinumTI " AutoEnc_gzistf46bgtcretjs26z2ktnuc ReEncodeFont " <[lbtn_gzistf.enc <fonts/HNSecWBiolinumTI.pfb}
\pdfmapline{=LinBiolinumTI-lf-t1--base HNSecWBiolinumTI " AutoEnc_pwsgbxvqjgn6vicsemw6brh2ee ReEncodeFont " <[lbtn_pwsgbx.enc <fonts/HNSecWBiolinumTI.pfb}
\pdfmapline{=LinBiolinumTI-lf-ts1--base HNSecWBiolinumTI " AutoEnc_5yotx4hr6bfsmiq44opstybx7d ReEncodeFont " <[lbtn_5yotx4.enc <fonts/HNSecWBiolinumTI.pfb}
\pdfmapline{=LinBiolinumTI-osf-ly1--base HNSecWBiolinumTI " AutoEnc_sakzps2xcrlkhximzvrta65xeh ReEncodeFont " <[lbtn_sakzps.enc <fonts/HNSecWBiolinumTI.pfb}
\pdfmapline{=LinBiolinumTI-osf-ot1--base HNSecWBiolinumTI " AutoEnc_vpjqewwvmxgjyq6acgkqjnibyb ReEncodeFont " <[lbtn_vpjqew.enc <fonts/HNSecWBiolinumTI.pfb}
\pdfmapline{=LinBiolinumTI-osf-sc-ly1--base HNSecWBiolinumTI " AutoEnc_7fko2hvq55cbusyotqz3zhzt7a ReEncodeFont " <[lbtn_7fko2h.enc <fonts/HNSecWBiolinumTI.pfb}
\pdfmapline{=LinBiolinumTI-osf-sc-ot1--base HNSecWBiolinumTI " AutoEnc_wruwpj6rvrvk5r47qngtvqeswd ReEncodeFont " <[lbtn_wruwpj.enc <fonts/HNSecWBiolinumTI.pfb}
\pdfmapline{=LinBiolinumTI-osf-sc-t1--base HNSecWBiolinumTI " AutoEnc_pagsaozuw2yzvy3etdhvrkbsac ReEncodeFont " <[lbtn_pagsao.enc <fonts/HNSecWBiolinumTI.pfb}
\pdfmapline{=LinBiolinumTI-osf-t1--base HNSecWBiolinumTI " AutoEnc_ouu7z6qpxnli2xfhbgmffeidab ReEncodeFont " <[lbtn_ouu7z6.enc <fonts/HNSecWBiolinumTI.pfb}
\pdfmapline{=LinBiolinumTI-osf-ts1--base HNSecWBiolinumTI " AutoEnc_5yotx4hr6bfsmiq44opstybx7d ReEncodeFont " <[lbtn_5yotx4.enc <fonts/HNSecWBiolinumTI.pfb}
\pdfmapline{=LinBiolinumTI-sup-ly1--base HNSecWBiolinumTI " AutoEnc_4gizrmphb34lvzu5433fj34dih ReEncodeFont " <[lbtn_4gizrm.enc <fonts/HNSecWBiolinumTI.pfb}
\pdfmapline{=LinBiolinumTI-sup-ot1 HNSecWBiolinumTI " AutoEnc_a6mubugl2s24lwap5jkwruu2la ReEncodeFont " <[lbtn_a6mubu.enc <fonts/HNSecWBiolinumTI.pfb}
\pdfmapline{=LinBiolinumTI-sup-t1--base HNSecWBiolinumTI " AutoEnc_3c7nkxqme2x5u3w3e6exxvmvkf ReEncodeFont " <[lbtn_3c7nkx.enc <fonts/HNSecWBiolinumTI.pfb}
\pdfmapline{=LinBiolinumTI-tlf-ly1--base HNSecWBiolinumTI " AutoEnc_n3xw57pbapmhxgbz6337fl5ynb ReEncodeFont " <[lbtn_n3xw57.enc <fonts/HNSecWBiolinumTI.pfb}
\pdfmapline{=LinBiolinumTI-tlf-ot1--base HNSecWBiolinumTI " AutoEnc_edcawa3wnysg7lbpamks34slvf ReEncodeFont " <[lbtn_edcawa.enc <fonts/HNSecWBiolinumTI.pfb}
\pdfmapline{=LinBiolinumTI-tlf-sc-ly1--base HNSecWBiolinumTI " AutoEnc_vezqth3hlczspa6x2yay6ze7qg ReEncodeFont " <[lbtn_vezqth.enc <fonts/HNSecWBiolinumTI.pfb}
\pdfmapline{=LinBiolinumTI-tlf-sc-ot1--base HNSecWBiolinumTI " AutoEnc_ep5ow4rox325nu34goxcx4eqkb ReEncodeFont " <[lbtn_ep5ow4.enc <fonts/HNSecWBiolinumTI.pfb}
\pdfmapline{=LinBiolinumTI-tlf-sc-t1--base HNSecWBiolinumTI " AutoEnc_jtta5hxbw7oyxjfmwnwi4i5o2a ReEncodeFont " <[lbtn_jtta5h.enc <fonts/HNSecWBiolinumTI.pfb}
\pdfmapline{=LinBiolinumTI-tlf-t1--base HNSecWBiolinumTI " AutoEnc_25tcsqj2ookahwk6t6j4gpln4d ReEncodeFont " <[lbtn_25tcsq.enc <fonts/HNSecWBiolinumTI.pfb}
\pdfmapline{=LinBiolinumTI-tlf-ts1--base HNSecWBiolinumTI " AutoEnc_afusaui7tlz4diouditdzd5ekd ReEncodeFont " <[lbtn_afusau.enc <fonts/HNSecWBiolinumTI.pfb}
\pdfmapline{=LinBiolinumTI-tosf-ly1--base HNSecWBiolinumTI " AutoEnc_c4mbatogxhrv24pydzfsgt5p2b ReEncodeFont " <[lbtn_c4mbat.enc <fonts/HNSecWBiolinumTI.pfb}
\pdfmapline{=LinBiolinumTI-tosf-ot1--base HNSecWBiolinumTI " AutoEnc_uqwtgwi22xmich6oxlj7e3ipwf ReEncodeFont " <[lbtn_uqwtgw.enc <fonts/HNSecWBiolinumTI.pfb}
\pdfmapline{=LinBiolinumTI-tosf-sc-ly1--base HNSecWBiolinumTI " AutoEnc_2or2lf6q5b7linlxmvytcuhdxd ReEncodeFont " <[lbtn_2or2lf.enc <fonts/HNSecWBiolinumTI.pfb}
\pdfmapline{=LinBiolinumTI-tosf-sc-ot1--base HNSecWBiolinumTI " AutoEnc_c5qhbuo23anriniajc5go5n2rd ReEncodeFont " <[lbtn_c5qhbu.enc <fonts/HNSecWBiolinumTI.pfb}
\pdfmapline{=LinBiolinumTI-tosf-sc-t1--base HNSecWBiolinumTI " AutoEnc_xeech75ax4jfjbpv6sdqtcqwvf ReEncodeFont " <[lbtn_xeech7.enc <fonts/HNSecWBiolinumTI.pfb}
\pdfmapline{=LinBiolinumTI-tosf-t1--base HNSecWBiolinumTI " AutoEnc_mx3chdhi3z7bhc5q3yql4n5qfc ReEncodeFont " <[lbtn_mx3chd.enc <fonts/HNSecWBiolinumTI.pfb}
\pdfmapline{=LinBiolinumTI-tosf-ts1--base HNSecWBiolinumTI " AutoEnc_afusaui7tlz4diouditdzd5ekd ReEncodeFont " <[lbtn_afusau.enc <fonts/HNSecWBiolinumTI.pfb}

%% file: Chapters/1_Introduction.tex
\section{Introduction}

Efficient approximate nearest neighbor (ANN) search~\cite{indyk1998approximate} over vector databases is a basic data analytical primitive and becomes a cornerstone of modern AI applications such as Retrieval-Augmented Generation (RAG)~\cite{lewis_retrieval-augmented_2020}, which enhances Large Language Models (LLMs) by incorporating external knowledge at inference time.
So far, designing an efficient ANN algorithm for a large-sized high-dimensional database has been well studied for decades~\cite{ann_survey_1, ann_survey_2, ann_survey_3}.

However, in real-world ANN applications, security and privacy concerns arise~\cite{ann_privacy_1, ann_privacy_2, ann_privacy_3}.
Without a protective protocol, either the querier submits the query in plaintext to the database owner, who executes the ANN algorithm, or the computation is delegated to the querier, which typically requires the database owner to reveal (at least part of) its data to enable search. The former violates query privacy, whereas the latter compromises database privacy. This motivates a symmetric private ANN model that simultaneously protects both the query and the database. We illustrate the need for symmetric privacy through two representative scenarios:

\textit{Example 1.1. In a joint healthcare study, a research institute \allowbreak(querier) prepares a drug trial by querying a hospital’s (database owner) private gene-expression database for similar patient profiles. The hospital returns only the top records allowed by the data-use agreement, and the institute uses them as LLM context to generate a trial plan. During the collaboration, disclosing the query may reveal the institute’s research details; disclosing the hospital’s database and patient data would violate regulations such as HIPAA.}

\textit{Example 1.2. During a cyber incident, a security team (querier) queries a vendor’s (database owner) private threat intelligence database for the closest matching attack reports and indicators, and uses the results to identify the threat and respond promptly. Disclosing the query can reveal what systems were hit; if the database is disclosed, it can expose sensitive intelligence sources and customer incidents.}

A naive approach to achieving symmetric privacy is to introduce a trusted third party, such as a cloud provider: both the database owner and the querier submit their plaintext vector database and queries to the third party, which then performs the ANN search and returns only the results to the querier. However, delegating all sensitive information in plaintext to a single, fully trusted entity may violate privacy regulations like GDPR and introduce severe risks of data breaches or misuse.

This motivates people to seek cryptographic solutions that can remove the need for trusted third parties.  One natural solution is to 
implement ANN search in the secure multi-party computation (MPC) model~\cite{FOCS:Yao86}.
In these solutions, the database owner and the querier secret-share the vector database and queries across multiple servers. The servers then jointly execute the ANN search protocol on these shares. The security of MPC ensures that no single server can reconstruct any input.
Several recent efforts pursue this direction. 
However, most existing solutions either relax privacy guarantees by assuming a public database~\cite{PacmannICLR25}, or incur high computational overhead~\cite{chen2020sanns,li2025panther}.
Specifically, the two closest MPC solutions to our setting~\cite{chen2020sanns,li2025panther} rely on cluster-based ANN pipelines: the database of size $N$ is partitioned into clusters, and the protocol identifies and searches the clusters most relevant to a query. Their efficiency depends heavily on the number of clusters, which is typically set to $\sqrt{N}$ to balance cluster selection and in-cluster search costs. As a result, their search time becomes $O(\sqrt{N})$ for each query. This quickly becomes impractical at modern database scales, because under MPC, each additional vector access incurs heavy communication, particularly for high-dimensional vectors. Earlier secure $k$NN works~\cite{songhori2015compacting, shaul2020secure, schoppmann2018private, zuber2021efficient} target smaller-scale or specialized two-party settings rather than scalable secret-shared ANN.

This motivates us to develop a more efficient MPC ANN protocol.
The natural idea is to deploy a state-of-the-art hierarchical graph-based ANN algorithm, such as HNSW~\cite{malkov2018efficient}, under the MPC setting.
By building a multi-layer graph index over the database, such algorithms enable efficient search: the query starts at the top layer, takes only a few steps per layer, and then moves to lower layers. It is widely accepted that, in practice, these indexes visit only $O(\log N)$ out of $N$ vectors per query in expectation, where $N$ is the database size.

However, implementing this hierarchical graph-based search in MPC is very challenging. 
At a high level, we must make two parts of the algorithm oblivious: (i) the search logic, e.g., when to stop the search, and (ii) the access pattern.
The stopping condition in HNSW is highly data- and query-dependent and replacing it with data-independent predicates without careful consideration totally compromises efficiency. 
For the access pattern, the search process may revisit the same node or vector multiple times, creating highly non-uniform and query-dependent memory accesses.
Some naive approaches to hide such redundant access patterns, such as ORAM~\cite{GolOst96}, would incur prohibitive communication overhead.
Therefore, this naturally raises a fundamental research question:

\subsection{Our Results}
\label{subsec:our_result}
In this paper, we give a positive answer to the above question by proposing the first MPC ANN protocol with logarithmic online query complexity.
Our protocol starts from the SOTA hierarchical graph index HNSW. 
We first study the single-query setting. 
To make the search logic fully oblivious, we design a data-independent control flow with a fixed search-termination bound and redesign all data structures as fixed-size arrays that support oblivious updates. 
To achieve an oblivious access pattern, we combine offline random reshuffling with a duplicated vector database, yielding $O(1)$ online overhead for each graph-index and vector access. Overall, our single-query MPC protocol achieves logarithmic online search complexity with at most a $2\times$ increase in vector storage. 
Moreover, we extend this design to support multiple queries. 
To handle repeated accesses across queries while avoiding additional online reshuffling, we introduce cross-query ``double visited'' structures that cache the graph nodes and vectors accessed by earlier queries, enabling later queries to reuse the cached information without revisiting the graph index and vector database.
In summary, our results are as follows:

\begin{enumerate}[wide=0pt,font={}]
  \item \textbf{Single query.}
  For a single query, \sysname{} achieves $O(\log N)$ online searching complexity over a database of size $N$.
  As Table~\ref{tab:ann_cost} shows, \sysname{} substantially outperforms state-of-the-art secure ANN baselines (SANNS~\cite{chen2020sanns}, PANTHER~\cite{li2025panther}), whose clustering-based pipelines have online communication that scales as $\sqrt{N}$.

  \item \textbf{Multiple queries.}
  For $U$ sequential queries, we propose a multi-query protocol with an amortized cost of  $O(U\log N)$.
  Unlike schemes requiring reshuffling per query (e.g., PANTHER~\cite{li2025panther}), our protocol allows $U$ queries to share a single offline reshuffle, avoiding additional online overhead.

  \item \textbf{Parallelism and batch queries.}
  Additionally, we design parallel implementations of the above protocols: for the single-query protocol, the critical-path length is $O(\log N)$, and for the multiple-query protocol, it is $O(U \log N)$. We also propose a batch-query protocol with higher parallelism that processes a batch of $U$ queries simultaneously with a much lower critical-path length of $O(\log^2 U + \log N)$.
  \item \textbf{Empirical evaluation.}
  We conduct a comprehensive experiment on large-scale ANN benchmarks. \sysname\ reaches recall over $0.9$, matching the state-of-the-art plaintext ANN algorithm. At the same target recall, \sysname\ achieves up to $22\times$ WAN speedup and $46\times$ less communication versus 2PC PANTHER; against the exact-top-$k$ 3PC SORT baseline, it achieves up to $1.6\mathrm{k}\times$ WAN speedup and $617\times$ less communication. All code is now available~\footnote{\url{https://github.com/Kleenex-ultrasoft/Secure_ANN}. Accessed: 2026-09-23.}.
\end{enumerate}

\begin{table}[!t]
\caption{\textbf{Online costs for top-$k$ ANN.}
\label{tab:ann_cost}
Communication and round complexity for one query and $U$ sequential queries over $N$ vectors of dimension $d$, with $k=O(1)$. $\tilde{O}(\cdot)$ hides polylogarithmic factors in $N$ and $d$.}
\centering
{\small
\setlength{\tabcolsep}{2pt}

\begin{tabular}{@{}>{\leavevmode\color{black}}l *{4}{>{\leavevmode\color{black}}c}@{}}
\toprule
& \multicolumn{2}{>{\leavevmode\color{black}}c}{\textbf{Single Query}}
& \multicolumn{2}{>{\leavevmode\color{black}}c}{\textbf{Multiple Queries}} \\
\cmidrule(lr){2-3}\cmidrule(lr){4-5}
\textbf{Scheme} & \textbf{Communication} & \textbf{Round} & \textbf{Communication} & \textbf{Round} \\
\midrule
SANNS~\cite{chen2020sanns} & $\tilde{O}(N^{1/2}d)$ & $\tilde{O}(1)$ & $\tilde{O}(U\,N^{1/2}d)$ & $\tilde{O}(U)$ \\
PANTHER~\cite{li2025panther} & $\tilde{O}(N^{1/2}d)$ & $\tilde{O}(1)$ & $\tilde{O}(U\,N^{1/2}d)$ & $\tilde{O}(U)$\\\midrule
HNSecW (Ours) & $\tilde{O}(d)$ & $\tilde{O}(1)$ & $\tilde{O}(Ud+U^2)$ & $\tilde{O}(U)$ \\
\bottomrule
\end{tabular}
}

\end{table}

%% file: Chapters/2_Preliminaries.tex
\section{Preliminaries}
\label{sec:preliminaries}

We introduce notation and background on ANN search, HNSW algorithm, secure multi-party computation, and oblivious RAM.

\subsection{Notation}
\label{subsec:notation}

Let $\mathcal{D} = \{\vec{v}_0, \vec{v}_1, \ldots, \vec{v}_{N-1}\}$ be a dataset of $N$ vectors in $\mathbb{R}^d$, where $d$ is the dimension. We use $\vec{v}[j]$ to denote the $j$-th coordinate of vector $\vec{v}$. Let $\mathit{dist}: \mathbb{R}^d \times \mathbb{R}^d \to \mathbb{R}_{\geq 0}$ denote a vector distance function. The index set is $[N] := \{0, 1, \ldots, N-1\}$. For any set $\mathcal{S}$, $|\mathcal{S}|$ is its cardinality. A graph $\mathcal{G} = (\mathcal{V}, \mathcal{E})$ has vertices $\mathcal{V} \subseteq \mathcal{D}$ and undirected edges $\mathcal{E}$. For any $\vec{v} \in \mathcal{V}$, $\mathcal{N}(\vec{v})$ denotes its neighbor set. 

\subsection{$k$-Nearest Neighbor Search}
\label{subsec:ann}

For the $k$-nearest neighbor ($k$-NN) problem, we have a dataset $\mathcal{D}$ of $N$ vectors in $\mathbb{R}^d$ and a query vector $\vec{q} \in \mathbb{R}^d$. The goal is to find the subset $\mathcal{D}^*_k(\vec{q}) \subseteq \mathcal{D}$ such that $|\mathcal{D}^*_k(\vec{q})| = k$, and for any $\vec{v} \in \mathcal{D}^*_k(\vec{q})$ and any $\vec{v}' \in \mathcal{D} \setminus \mathcal{D}^*_k(\vec{q})$, it holds that $\mathit{dist}(\vec{v}, \vec{q}) \leq \mathit{dist}(\vec{v}', \vec{q})$. A straightforward approach is the linear scan, which computes the distance from $\vec{q}$ to all $N$ vectors and maintains the $k$ smallest distances found so far, typically using a priority queue of size $k$.  The computational cost is $O(Nd + N \log k)$, prohibitive for large-scale datasets (large $N$) or high-dimensional data (large $d$).

To address the $O(Nd + N \log k)$ cost, graph-based $k$-NN approaches have been proposed. These methods operate on a graph $\mathcal{G} = (\mathcal{D}, \mathcal{E})$ over the dataset $\mathcal{D}$, where each vector in $\mathbb{R}^d$ is represented as a vertex, and edges connect pairs of sufficiently ``close'' vectors. Given a query $\vec{q}$, the search greedily traverses the graph from an initial node toward nodes closer to $\vec{q}$. A representative example is the Delaunay graph~\cite{dobkin1990delaunay}. In the two-dimensional case, each node in the Delaunay graph has an expected degree of $O(1)$, and it guarantees that for any query $\vec{q}$, the exact nearest neighbor can be found in expected $O(\log N)$ time~\cite{malkov2018efficient}. However, moving towards the general case, the average node degree in the Delaunay graph grows exponentially in vector dimension with $O(N^{\lceil d/2 \rceil})$~\cite{dobkin1990delaunay}, making search impractical.

\subsection{Approximate $k$-NN and Hierarchical Navigable Small World}
\label{subsec:hnsw}

\begin{algorithm}[!t]
\caption{\textsc{SearchLayer}: Greedy search in HNSW layer}
\label{alg:search-layer}
\small
\begin{algorithmic}[1]
\Require Query $\vec{q}$, enter point $\mathit{ep}$, search width $\mathit{ef}$, layer $\ell$
\Ensure Set $W$ of $\mathit{ef}$ closest neighbors to $\vec{q}$ in layer $\ell$

\State $V \gets \{\mathit{ep}\}$ \Comment{Visited set}
\State $C \gets \{\mathit{ep}\}$ \Comment{Candidate set}
\State $W \gets \{\mathit{ep}\}$ \Comment{Result set}

\While{$|C| > 0$}
    \State $c \gets$ extract nearest element from $C$ to $\vec{q}$
    \State $f \gets$ furthest element from $W$ to $\vec{q}$

    \If{$\mathit{dist}(c, \vec{q}) > \mathit{dist}(f, \vec{q})$}
        \State \textbf{break} \Comment{Cannot improve $W$ further}
    \EndIf

    \For{each $e \in \mathcal{N}_\ell(c)$}
        \If{$e \notin V$}
            \State $V \gets V \cup \{e\}$
            \State $f \gets$ furthest element from $W$ to $\vec{q}$

            \If{$\mathit{dist}(e, \vec{q}) < \mathit{dist}(f, \vec{q})$ \textbf{or} $|W| < \mathit{ef}$}
                \State $C \gets C \cup \{e\}$
                \State $W \gets W \cup \{e\}$

                \If{$|W| > \mathit{ef}$}
                    \State remove furthest element from $W$ to $\vec{q}$
                \EndIf
            \EndIf
        \EndIf
    \EndFor
\EndWhile

\State \Return $W$
\end{algorithmic}
\end{algorithm}

Given that existing $k$-NN search algorithms are infeasible for large-scale, high-dimensional data, where both linear scan and ideal graph-based methods are prohibitively expensive, a practical trade-off between accuracy and efficiency is necessary. Consequently, people have proposed the \textit{Approximate $k$-NN (ANN)}. In ANN, instead of returning the exact nearest set $\mathcal{D}^*_k(\vec{q})$, the goal is to find an approximate set $\mathcal{S}$ that meets a target Recall $R \in (0,1]$, defined as $\operatorname{Recall}(\mathcal{S}) = |\mathcal{S} \cap \mathcal{D}_k^*(\vec{q})| / k \geq R$.
Over the years, various families of ANN methods have been developed, including hashing-based, tree-based, and quantization-based approaches. Among them, Hierarchical Navigable Small World (HNSW)~\cite{malkov2018efficient}, achieves state-of-the-art query performance for high-dimensional data~\cite{malkov2018efficient}.

The core of HNSW is a multi-layered graph structure $\mathcal{G} = \{\mathcal{G}_0, \mathcal{G}_1, \ldots, \mathcal{G}_{L-1}\}$, where each layer $\mathcal{G}_\ell = (\mathcal{V}_\ell, \mathcal{E}_\ell)$ denotes a graph whose nodes correspond to vectors in the dataset $\mathcal{D}$. The layers form a sequence of nested subsets of $\mathcal{D}$: $\mathcal{V}_{L-1} \subset \mathcal{V}_{L-2} \subset \cdots \subset \mathcal{V}_1 \subset \mathcal{V}_0 = \mathcal{D}$.
The number of nodes decreases geometrically from bottom to top, resulting in $L = O(\log N)$ layers. 
The edges are constructed from the nodes' underlying vector representations, and each $\mathcal{G}_\ell$ serves as an out-degree bound $M=O(1)$ approximation of a Delaunay graph.
This hierarchical graph $\mathcal{G}$ is an efficient index for the vector database $\mathcal{D}$. Its design enables a two-phase search strategy: upper, sparser layers support coarse-grained navigation, ensuring high search efficiency, while lower, denser layers provide fine-grained refinement, ensuring high search accuracy.

For query $\vec{q} \in \mathbb{R}^d$, HNSW starts at a random entry point in the top layer $\mathcal{G}_{L-1}$.
Each layer search locates the locally closest node to $\vec{q}$, which becomes the entry point for the next lower layer. 
After searching the base layer $\mathcal{G}_0$, the algorithm returns the $k$ nearest neighbors of $\vec{q}$ from all visited nodes.
To implement this idea, at each layer the search maintains three data structures: a candidate set $C$, which stores potential nodes for further exploration; a result set $W$, which stores the closest nodes found so far, where $|W| = \mathit{ef}$ with $\mathit{ef}=1$ for non-base layers and $\mathit{ef}=O(k)$ for the base layer; and a visited set $V$, which records all visited nodes to prevent revisits. 

All these data structures are initialized with the entry point $\mathit{ep}$, and the search then proceeds iteratively. At each iteration, we extract the nearest candidate $c$ from $C$, explore each unvisited neighbor $e$ of $c$, and add them to $V$. 
If $e$ is closer to $\vec{q}$ than the farthest node in $W$, it is inserted into both $C$ and $W$, and $W$ is kept at size at most $\mathit{ef}$. The search terminates when no candidate in $C$ is closer to $\vec{q}$ than the farthest node in $W$, or when $C$ becomes empty. Algorithm~\ref{alg:search-layer} gives the detailed procedure.

We abstract HNSW into three key components: the \textbf{search logic flow}, the \textbf{data structures}, and the \textbf{graph and vector access}. The logic flow specifies how the search proceeds within each layer. In each step, it moves to the nearest candidate node and then explores all its neighbors. The search stops when no candidate is closer to the query than the current results. The data structures support this process. At each layer, the algorithm maintains three sets: a candidate set $C$, a result set $W$, and a visited set $V$. 
Finally, the logic flow interacts with the underlying graph index and vector database. For each candidate $c$, the search first accesses the graph index to retrieve its neighbor list, then accesses the vector database to obtain these neighbors' vectors for distance computation.

\subsection{Secure Multi-Party Computation}
\label{subsec:mpc}
Secure multi-party computation (MPC) enables multiple parties to jointly compute a function over their private inputs while only revealing the output to designated parties~\cite{FOCS:Yao86}. 
We use secret-sharing MPC protocols where, for each intermediate value $x$, all parties hold its secret sharing denoted as $\llbracket x \rrbracket$.
The security ensures that the adversary cannot learn the value even when corrupting some number of parties as allowed by the threat model.
Our system treats MPC protocols as black boxes; thus, we describe computation directly over secret shares.
In particular, we use
$\share{x}+\share{y}$ and 
$\share{x}\times\share{y}$
to denote 
addition and multiplication
in MPC over secret-shared values. 
Typically, given network bandwidth $\mathsf{BW}$ and round-trip time $\mathsf{RTT}$, MPC protocols follow this online latency model:
\begingroup
\[
T_{\mathrm{online}} \approx \mathsf{Round}\cdot \mathsf{RTT} \;+\; \frac{\mathsf{Communication}}{\mathsf{BW}} \;+\; \mathsf{Computation}.
\]
\endgroup
Here, Computation is each party’s local, non-interactive work and is often almost free. Communication is the total data exchanged and scales linearly with the multiplication gate count. Round is the number of sequential interaction steps, determined by the circuit’s multiplicative depth.

\noindent\textbf{Making Algorithms Oblivious.}
Under MPC, oblivious execution has three requirements:
\begin{enumerate}[label=(\arabic*),wide=0pt,font={}]
\item \textbf{Logic flow.} For fixed public parameters, the algorithm follows the same sequence of operations. For example, each loop runs for a fixed number of iterations.
\item \textbf{Data structures.} For fixed public parameters, intermediate data structures have the same sizes and update schedule.
\item \textbf{Data access pattern.} For fixed public parameters, physical-address sequences are indistinguishable across private inputs.
\end{enumerate}

\noindent {\bf Oblivious Access in MPC.}
Among these requirements, oblivious data access has been widely studied as a general primitive.
MPC protocols based on secret-sharing can typically support any circuit, but random accesses are expensive to represent as circuits. To support private accesses on large arrays without incurring a cost linear to the array, one would need to use an Oblivious RAM (ORAM)~\cite{GolOst96} by executing the ORAM client in MPC~\cite{CCS:GKKKMR12}.
This protocol is also called Distributed ORAM (DORAM).
Given a secret-shared array $\llbracket \vec{v} \rrbracket$ and secret-shared index $\llbracket i \rrbracket$, DORAM enables parties to jointly compute $\llbracket \vec{v}[\llbracket i \rrbracket] \rrbracket$ (the secret-shared value at secret-shared position) without revealing the index or the value to any party.
DORAM has been optimized in different ways, including circuit-based minimization (e.g.,~\cite{CCS:WanChaShi15}) and protocol-level adaptation (e.g.,~\cite{doerner2017scaling}).
Although the most efficient construction can achieve $O(\log N)$ communication overhead~\cite{vadapalli2023duoram}, DORAM still incurs an $O(N)$ computational cost.
Moreover, DORAM accesses are stateful and often sequential, making batched accesses difficult to parallelize.
Therefore, rather than directly adopting generic DORAM in our MPC protocol, we propose a specialized algorithm to implement an oblivious access pattern.

\noindent\textbf{Security Models.}
Beyond these requirements, oblivious algorithm design also depends on the security model. For secure ANN, we distinguish four models based on who has plaintext access to each input and who performs the search, from general to specific:
\begin{enumerate}[label=(\arabic*),leftmargin=*,font={}]
\item \textbf{Symmetric MPC.} The computation parties have no access to the database, index, or query in plaintext. They perform the search over secret values. This is also the standard setting in the MPC literature. Our paper targets this model for data collaboration between untrusted organizations and computation outsourcing by a single data owner.
\item \textbf{Client-server search.} The client has plaintext access to the query, while the server has plaintext access to the database and index. They jointly perform the search without learning the other's input beyond the result. SANNS and PANTHER follow this model~\cite{chen2020sanns,li2025panther}.
\item \textbf{Storage outsourcing.} The client has plaintext access to the database, index, and query. It encrypts the database and index and then outsources them to an untrusted server. During search, the client retrieves data from the server and performs the search locally. Compass follows this model~\cite{zhu_compass_2025}.
\item \textbf{Query-private search.} The client has plaintext access to the query. The database and index are public and stored on the server, so both the client and server have plaintext access to them. During search, the client retrieves database and index entries from the server and searches locally. Thus, only the query is hidden from the server. Pacmann follows this model~\cite{PacmannICLR25}.
\end{enumerate}

\begin{figure}[!t]
    \centering
    \papergraphics[width=0.99\linewidth,trim=0 6.11bp 0 25.00bp,clip]{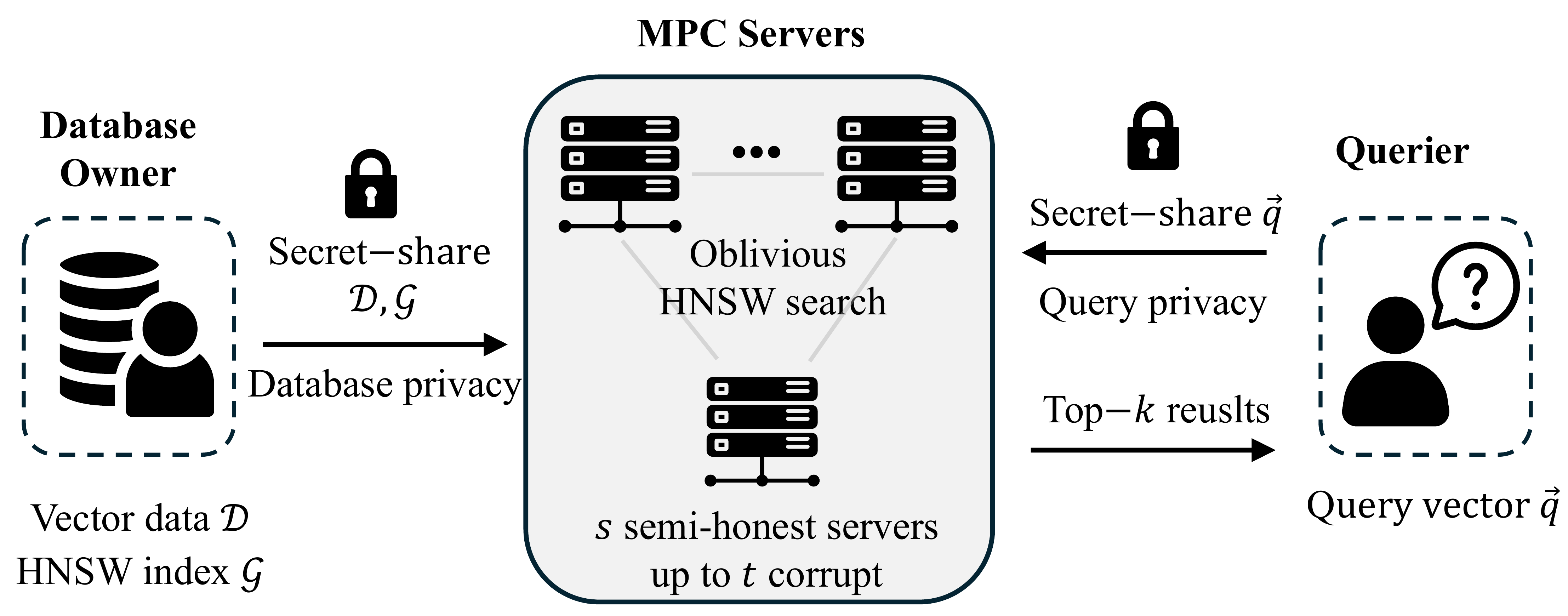}
    \caption{System overview of Oblivious HNSW}
    \label{fig:system_overview}
\end{figure}

%% file: Chapters/3_System_Overview.tex
\section{System Overview}

We consider a private HNSW search involving two input-holding parties and a set of $s$ MPC servers; Figure~\ref{fig:system_overview} gives the high-level architecture.
The two input parties are: (1) the database owner (DO), who holds a private vector dataset $\mathcal{D} = \{\vec{x}_i\}_{i=0}^{N-1} \subset \mathbb{R}^d$ together with an HNSW graph index $\mathcal{G}$ built over $\mathcal{D}$, and
(2) the Querier, who holds the private query vector $\vec{q} \in \mathbb{R}^d$.
DO and Querier are two logical parties who do not need to trust each other, but they could be instantiated as the same physical party, in which case the protocol reduces to the standard outsourced computation model.
The functionality of the protocol is shown in Figure~\ref{func}.

\begin{figure}[!t]
  \setlength{\FrameSep}{4pt}
  {\small
  
  \begin{nffunc}{$\Func[HNSW][\QQQ]$}
    \raggedright
    \noindent{\bf Public Input:} $\QQQ : (\mathbb{R}^d)^N \times \GGG \times \mathbb{R}^d \rightarrow (\mathbb{R}^d)^k$.
    
    \noindent{\bf Inputs:} $\mathcal{D} = \{\vec{x}_i\}_{i=0}^{N-1} \subset \mathbb{R}^d$, $\vec{q} \in \mathbb{R}^d$, and $\mathcal{G}$

    \smallskip\noindent{\bf Compute:} Upon receiving $({\sf dataset}, \mathcal{D}, \mathcal{G})$ from DO, and $({\sf query}, q)$ from Querier, where $\mathcal{G}$ is an HNSW graph,
            compute $(\vec{v}_0,\ldots,\vec{v}_{k-1}) \leftarrow \QQQ(\mathcal{D},\mathcal{G},\vec{q})$

    \noindent{\bf Outputs:}
    Upon receiving $({\sf continue})$ from all servers,
    send $(\vec{v}_0,\ldots,\vec{v}_{k-1})$ to $\Querier$.

  \end{nffunc}
  }
  \caption{Ideal functionality for securely computing HNSW.}
  \label{func}
\end{figure}

\subsection{Threat Model}
\label{subsec:threat}

We assume a semi-honest, probabilistic polynomial-time (PPT) adversary who may corrupt at most $t$ MPC servers and one or both input parties.
Our protocol achieves standard simulation-based security~\cite{JC:Canetti00}.
For any PPT adversary corrupting at most $t$ MPC servers and one or both input parties, there exists a PPT simulator whose output is computationally indistinguishable from the adversary’s output, using only the corrupted parties’ inputs, their outputs under $\mathcal{F}_{\mathrm{HNSW}}$, and the public parameters. We provide the full proofs in the extended version~\cite{hnsecw_extended}.
Table~\ref{tab:public_parameters} summarizes the public parameters.

\begin{table}[!t]
\caption{Public parameters of the system.}
\centering
{\small

\begin{tabularx}{\linewidth}{@{}>{\leavevmode\color{black}}l >{\leavevmode\color{black}}X@{}}
\toprule
\textbf{Category} & \textbf{Public Parameters} \\
\midrule
Data characteristics 
  & Vector dimension $d$; Dataset size $N$; Number of layers $L$; Degree bound $M$; Per-layer sizes $\{N_\ell\}^{L-1}_{\ell=0}$ \\
Search hyperparameters 
  & Result set size $\mathit{ef}$; Nearest neighbor output size $k$ \\
Multi/Batch-query setting 
  & Number of queries $U$ \\
\bottomrule
\end{tabularx}
}

\label{tab:public_parameters}
\end{table}

The ideal functionality and the threat model imply symmetric privacy and correctness.
For \textbf{query privacy}, the database owner and all MPC servers learn nothing about $\vec{q}$ beyond what is implied by the public parameters.
For \textbf{database privacy}, the querier and all MPC servers learn nothing about $\mathcal{D}$ or $\mathcal{G}$ beyond the top-$k$ result vectors output by $\mathcal{F}_{\mathrm{HNSW}}$ and what is implied by the public parameters.
For \textbf{correctness}, for any inputs $(\mathcal{D},\mathcal{G},\vec{q})$, the Querier reconstructs the output of the ideal functionality $\mathcal{F}_{\mathrm{HNSW}}$.

\subsection{Single-Query}
Even when all vectors and their distance to the query are kept secret, a single HNSW query is not oblivious in its control flow, data structures, or graph and vector access patterns, all depending on the underlying graph and query. 
Achieving obliviousness to prevent leakage requires a dedicated analysis of the performance features that HNSW presents, since a generic method that follows only worst-case conditions would introduce prohibitive overhead.
In our protocol, we design customized optimizations to make each part oblivious by taking advantage of its unique features.
For a single query, we traverse each layer without revealing the access pattern. At every layer, we use our oblivious data structure to maintain the search state, and we terminate the traversal using a tailored termination condition that prevents information leakage. 
We outline the key ideas here, with full details in Section~\ref{subsec:single_query}.
\begin{itemize}[wide=0pt,font={}]
    \item {\bf Control flow.}
    We eliminate data-dependent stopping conditions by replacing them with a fixed iteration bound derived from a detailed analysis of HNSW’s stopping behavior, ensuring obliviousness with only negligible recall loss (Section~\ref{subsec:logic_oblivious}). Although there is no theoretical bound on the iterations, we found that in the specific context of RAG, we could set a tight upper bound on the number of iterations without degrading the end-to-end performance.

    \item {\bf Data structure.} We observe that both the candidate set and the result set exhibit a monotonic exclusion property: once a node is excluded, it will not be included in these sets again. Using this monotonicity, we design an oblivious data structure for both sets that keeps them at a fixed size and removes farther nodes through fixed-size tail truncation after oblivious sorting (Section~\ref{subsec:structure_oblivious}).
    \item {\bf Graph and vector access pattern.} We use the fact that HNSW naturally avoids revisiting a node more than once per layer.
    To further reduce redundant access across the layers, we assign each layer an independently shuffled graph and vector tables so every address is accessed at most once, yielding fully oblivious access patterns.
\end{itemize}

\subsection{Multiple-Query}
Extending the single-query protocol to multiple queries introduces a fundamentally new challenge: different queries may revisit the same graph node and its associated vector. Hiding these repeated visits would require reshuffling the graph and vector tables after every query. However, such per-query reshuffling is impractical in an online setting due to the significant latency it entails. 
To address this, we rely on the insight that once data has been accessed, it can be cached and reused across queries rather than accessed repeatedly. We propose to use two global cross-query visited structures, one for graph neighbor lists and one for vector contents. These structures serve as a secure shared cache: once any query retrieves a node or vector, it is stored in the global visited array. Subsequent queries check this array first and reuse the cached data. 
As a result, each graph or vector access first scans the cache and then performs a real or dummy data-access step. Under this design, every node and vector is accessed at most once from the underlying shuffled tables globally, while subsequent revisits are served through oblivious cache scans. Combined with one-time random pre-shuffling, this keeps the overall access pattern oblivious; see Section~\ref{sec:multi}.

\subsection{Parallel-Enhanced Batch Model}

Furthermore, we consider how the oblivious HNSW search can be executed in parallel. The single-query protocol already exhibits high parallelism, but its multiple-query extension is designed to process the $U$ queries sequentially, giving a linear circuit depth of $O(U)$. In a batch setting where many queries are available simultaneously, it is natural to ask whether the computation can be reorganized to reduce this depth to better exploit parallel execution.
As in the multiple-query setting, the fundamental challenge is to avoid accesses to the same nodes and vectors across queries. 
A naive approach is to cache every accessed element (neighbor lists and vectors), as in the sequential protocol. 
But in the batch setting, such a large cache is unnecessary. Our key insight is that, because all queries proceed in synchronized iterations, we do not passively wait for cache hits and reuse; instead, we force $U\cdot M$ accessed vectors to be evaluated against $U$ queries: if a query later needs that vector, it can skip the inner path; if not, the extra evaluation does not affect correctness. 
This eliminates the need for a massive cache and naturally prevents redundant accesses. 
As a result, the batch-query protocol achieves a depth of $O(\log^2 U + \log N)$, coming from the oblivious duplicate removal of the $U \cdot M$ neighbors per iteration plus the $O(\log N)$ HNSW layers.

\subsection{Implementation \& Experiment Summary}

We implement \sysname\ in two MPC configurations: 2PC in ABY \cite{demmler2015aby} and 3PC in MP-SPDZ~\cite{keller2020mp}. We evaluate on $10$ ANN benchmarks at $M{=}128$ (SIFT-1M, GIST-1M, DEEP-100M, BIGANN-1B, DEEP-1M, DEEP-image, MNIST-60K, Fashion-60K, SPACEV-1B, SSNPP-1B)~\cite{jegou2010product,babenko2016efficient,simhadri2022bigann,lecun1998gradient,xiao2017fashion}, spanning $N$ from $60K$ to $1B$ and $d{\in}[96,960]$. Experiments are conducted under LAN ($4000$~Mbps, $1$~ms RTT) and WAN ($320$~Mbps, $50$~ms RTT).
On accuracy, \sysname\ delivers Recall@$k\ge 0.9$ for at least $95\%$ of queries across the full evaluated range of $k\in\{10,20,50,100\}$. On performance, \sysname\ outperforms existing work on WAN 2PC by up to $22{\times}$ in end-to-end latency and $46{\times}$ in communication.

%% file: Chapters/4_Single_Query.tex
\section{Single Query Protocol}
\label{subsec:single_query}

As discussed in Section~\ref{sec:preliminaries}, we decompose HNSW search into three components: the search logic flow, the data structures, and the graph/vector access patterns. In this section, we briefly outline how to make each component oblivious.
Section~\ref{subsec:logic_oblivious} makes the control flow data-independent by replacing the original data-dependent stopping conditions with a fixed iteration bound. Section~\ref{subsec:structure_oblivious} redesigns the visited, candidate, and result structures as fixed-size arrays with uniform update rules. Section~\ref{subsec:access_oblivious} ensures oblivious graph and vector access by avoiding repeated retrievals through per-layer reshuffled tables. Finally, Section~\ref{sec:put_it} assembles these building blocks into the complete oblivious single-query protocol. 

\subsection{Making Logic Flow Oblivious}
\label{subsec:logic_oblivious}
We first make HNSW's search logic flow oblivious. At each layer, original HNSW iteratively finds the nearest candidate node and explores all its neighbors. The logic flow is inherently data-dependent.

The first source of data dependence is varying node degrees: exposing this variation would leak information about the graph and query. Fortunately, this dependency is easy to remove. 
Since HNSW enforces an out-degree upper bound $M$ per node, we pad each node's neighbor list with randomly selected nodes from the same layer to obtain a fixed length of $M$.
This padding does not reduce search quality: all real neighbors remain available for selection.

The more subtle challenge lies in the termination condition. In HNSW, the search stops when no better candidate remains (line 7-8 in Algorithm \ref{alg:search-layer}), or when the candidate set becomes empty (line 4 in Algorithm \ref{alg:search-layer}). Both conditions are inherently data-dependent.
At first glance, this seems similar to the bounded-degree case, since the original HNSW paper~\cite{malkov2018efficient} suggests that each layer can be traversed in a constant number of hops. One may use that constant as a fixed upper bound and pad execution with dummy searches. However, unlike node degree, this constant is not a worst-case bound; it arises from an expected-case analysis in~\cite{malkov2018efficient} that assumes each layer forms a perfect Delaunay graph~\cite{dobkin1990delaunay}. Real HNSW layers only approximate such structures, so the constant-hop guarantee does not hold in general.
In fact, the best-known asymptotic bound for HNSW on arbitrary data is still $O(N)$.  A practical, data-independent iteration bound must therefore be derived from the actual termination behavior.

The first termination condition (lines~7--8 in Algorithm~\ref{alg:search-layer}) triggers when the closest candidate can no longer improve the current result set. 
This can only happen after $\mathit{ef}$ search steps. 
Since every node inserted into $W$ is also inserted into $C$, every node in the final $W$ must have been popped from $C$ and explored (i.e., cannot remain in $C$); otherwise, it would stay in $C$ and lead to $\min(C)\le \max(W)$, contradicting the stopping condition.
Hence, letting $s$ denote the number of search steps performed by standard HNSW at a given layer, we always have $s\ge \mathit{ef}$ and can write exact search steps as $s = \mathit{ef} + \Delta$, where residual $\Delta$ captures work beyond the $\mathit{ef}$ steps.  
Therefore, to make the termination condition oblivious, a natural approach is to fix the number of search steps at each layer. 
Specifically, for each layer $\ell$, we set the number of search steps to $T^{(\ell)} = \mathit{ef}^{(\ell)} + \overline{\Delta}^{(\ell)}$, where $\overline{\Delta}^{(\ell)}$ is a public upper parameter to bound the residual term and depends only on public quantities such as $N$, $d$, and $M$.
Below, we present our strategy for setting $\mathit{ef}^{(\ell)}$ and $\overline{\Delta}^{(\ell)}$, both bounded by logarithmic terms, and show the resulting procedure performs robustly in practice.
\emph{However, designing termination strategies for HNSW that balance accuracy and efficiency remains an active research topic even in the non-private setting~\cite{chatzakis2025darth}, and is not the main interest of this paper. Future termination strategies can also be integrated into our framework. Moreover, we highlight that, as long as the number of search steps increases by at most a logarithmic factor, our final complexity guarantee remains unchanged.}

\begin{figure}[!t]
    \centering
    \papergraphics[width=\linewidth,trim=0 2.67bp 0 1.50bp,clip]{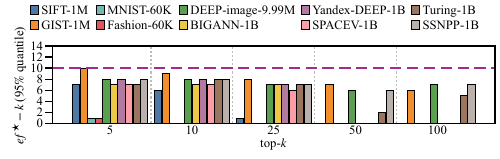}
    \caption{Base-layer $\mathit{ef}^*$ on the $10$ datasets ($M{=}128$) with $95\%$ recall $\ge 0.9$ over $10^3$ queries.}
    \Description{Grouped bar chart of ef^*-k versus top-k for 10 ANN datasets.}
    \label{fig:ef9ds-calib}
\end{figure}

We begin with the choice of $\mathit{ef}^{(\ell)}$. Standard HNSW~\cite{malkov2018efficient} sets $\mathit{ef}^{(\ell)}=1$ in all upper (non-base) layers, while treating the base-layer value as an input parameter.  We now discuss how to set $\mathit{ef}^{(0)}$ at the base layer.
To this end, we run HNSW with five values of $k$ on $10$ ANN datasets spanning $60$K--$1$B points and dimensions $96$--$960$, using $10^3$ queries per dataset. For each test, let $\mathit{ef}^*$ denote the smallest base-layer value of $\mathit{ef}^{(0)}$ that achieves recall at least $90\%$. Figure~\ref{fig:ef9ds-calib} reports the $95\%$ quantile of $\mathit{ef}^*$ over these $50$K tests. For all $k \ge 5$, the offset $\mathit{ef}^* - k$ is at most $10$ in at least $95\%$ of the tests. Therefore, we set $\mathit{ef}^{(0)}=k+10$ throughout our experiments, except in the ablation study on $\mathit{ef}$.

\begin{figure}[!t]
    \centering
    \papergraphics[width=\linewidth,trim=0 1.90bp 0 0.50bp,clip]{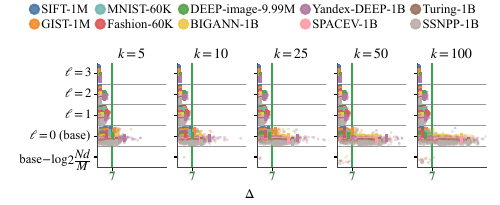}
    \caption{Per-layer residual $\Delta$ at $M{=}128$, $\mathit{ef}^{(\ell\geq 1)}{=}1$, $\mathit{ef}^{(0)}{=}k{+}10$, on the $10$ datasets. The bottom row of each panel shows the base-layer step count minus $\lceil\log_2(Nd/M)\rceil$.}
    \Description{Per-layer residual scatter plot across 10 datasets at five top-k values.}
    \label{fig:per-layer-delta}
\end{figure}

After fixing $\mathit{ef}^{(\ell)}$, we discuss how to set $\overline{\Delta}^{(\ell)}$. One natural idea is to set $\overline{\Delta}^{(\ell)}$ to one empirical constant across all layers. We validate this strategy in the same $50$K experiments by counting the residual steps required by standard HNSW under the above strategy for $\mathit{ef}^{(\ell)}$. Figure~\ref{fig:per-layer-delta} shows the results.

The results show that the residual steps are well bounded by a constant in the upper layers, but are substantially larger at the base layer.
We therefore use an enhanced strategy. For upper layers, we set $\overline{\Delta}^{(\ell)}$ to a constant $C$. For the base layer, we use the observation from recent work on early termination in HNSW~\cite{chatzakis2025darth,chatzakis2026darthplus} that, after the first $\mathit{ef}$ steps, the probability of not having terminated decays geometrically with the number of additional steps. We therefore add a logarithmic term in $N$, $d$, and $M$ at the base layer: $\overline{\Delta}^{(0)}=\left\lceil \log_2(Nd/M) \right\rceil + C$.
 In our experiments, we set $C=7$, which
covers more than $95\%$ of cases across the $50$K tests.

We therefore apply this fixed-step rule: $1+7=8$ search steps at each upper layer and $\lceil\log_2(N\!\cdot\!d/M)\rceil+\mathit{ef}^{(0)}+7$ at the base layer.
Since $\overline{\Delta}^{(\ell)}$ may be larger than the actual number of residual steps, and since the second termination condition in line~4 of Algorithm~\ref{alg:search-layer} may trigger early termination, the search may otherwise stop before completing $\overline{\Delta}^{(\ell)}$ steps. In this case, we continue the search using dummy nodes. When dummy nodes are treated as candidates during the search, they are assigned sentinel distances that rank worse than all real candidates, and thus never affect the result set $W$.
Overall, our strategy yields an oblivious search procedure.
Here, the number of search steps is $O(1)$ at each upper layer and $O\bigl(\log(N \cdot d/M)\bigr)$ at the base layer. Since HNSW has $O(\log N)$ layers, this gives a total of
$O\bigl(\log(N \cdot d/M) + \log N\bigr)$ search steps per query.

\subsection{Making Data Structures Oblivious}
\label{subsec:structure_oblivious}

We next make the data structures of HNSW oblivious.
Recall that each layer individually maintains three structures: the visited set $V$, the candidate set $C$, and the result set $W$. 
However, the sizes and update frequencies of the three sets are inherently data- and query-dependent during search. 
We therefore redesign them into oblivious counterparts.

\subsubsection{Oblivious Result Set $W$}
The result set $W$ is easier to make oblivious than $C$ and $V$ because original HNSW already bounds its size by $\mathit{ef}$.
The only data-dependent behavior is the conditional insertion rule, i.e., inserting a node into $W$ when it improves the current $\mathit{ef}$ results and passes the visited check (line 13 in Algorithm \ref{alg:search-layer}). 
We remove this data-dependent condition by obliviously considering every node for insertion, at a cost of $O(\mathit{ef})$ per node.\footnote{One could reduce this to $O({\sf polylog}(\mathit{ef}))$ using more sophisticated oblivious algorithms~\cite{asharov2022efficient}, but these become advantageous only when $\mathit{ef}$ is very large ($\geq 10^4$), which is far beyond our setting.} Overall, this naive approach costs $O(M \cdot \mathit{ef})$ in each iteration.
For practical efficiency, we instead insert all $M$ elements, apply oblivious sorting, and then reduce the size back to $\mathit{ef}$, incurring only $O(\mathit{ef} \log^2(\mathit{ef}))$ complexity. Since $\log^2(\mathit{ef})$ is much smaller than $M$ in our setting, this approach outperforms the naive alternative in practice.

\subsubsection{Oblivious Visited Set $V$}
\label{subsubsec:oblivious_visited}
The visited set $V$ records explored nodes to prevent later revisits. 
In the original HNSW, $V$ is a dynamic set whose size and insertion pattern are both query- and data-dependent, since nodes are inserted only when first encountered and their growth is unpredictable. 
To eliminate this dependence, we observe that $V$ can safely contain duplicate nodes. 
We therefore replace the set, whose size is unpredictable, with a list whose size increases by exactly one after each exploration.
More precisely, each iteration appends all $M$ neighbors, regardless of whether they have appeared before, so that $V$ contains exactly $M\cdot T + 1$ entries after $T$ iterations.
A full linear scan of this list performs a visited check at a cost of $O(MT)$, which is scalable and independent of $N$.

\subsubsection{Oblivious Candidate Set $C$}
The candidate set $C$ stores the nodes awaiting expansion and is also a fully dynamic set.
In the original HNSW, its size is inherently data-dependent: each iteration extracts one node from the set but may insert zero, a few, or up to $M$ new nodes.
The insertion condition is also data-dependent, since neighbors are added to $C$ only when they pass the visited check and could improve the result set $W$ (line 10, 13 in Algorithm \ref{alg:search-layer}).
To make $C$ oblivious, we leverage the fixed iteration $T$ established in
Section~\ref{subsec:logic_oblivious}.
Since at most $T$ candidates will be extracted, entries beyond the top-$T$ positions cannot be used and need not be retained.
We therefore replace the dynamic set with a fixed-length array of size $T$.
In every iteration, all $M$ neighbors, whether real or dummy after the visited check, are appended to $C$. We then perform an oblivious sort by distance and retain only the top-$T$ entries. 
As we discussed earlier, since dummy nodes carry a maximum distance $d_{\text{max}}$, they always expand to other dummy nodes. They cannot influence the result set $W$ even if they are selected after the real nodes are exhausted in $C$.
This fixed-size and fixed-update design of $C$ eliminates data-dependent behavior while preserving original HNSW's functionality.

\subsection{Making Access Patterns Oblivious}
\label{subsec:access_oblivious}

The HNSW search process involves two types of data access: vector access and graph access.
During the search, a node first incurs a vector access where its distance to the query is evaluated. If the node is later selected for expansion (extracted from the candidate list $C$), it then incurs a graph access to retrieve its neighbor list. A straightforward way to make both accesses oblivious is to store the graphs and the vector database in DORAM. However, this incurs significant overhead, as each access requires $O(\log N_{\ell})$ time.

To design a more efficient solution, we recall that achieving oblivious access requires two properties: data anonymization, which can be easily achieved through pre-shuffling, and non-repetitive access, where no address is revisited within a query. In this single-query setting, we assume that both graph and vector data have been anonymized through offline reshuffling. Therefore, if each item is retrieved at most once, the access is oblivious. We now analyze where repeated accesses arise and how to avoid them.

Within a single layer, both graph and vector accesses are non-repetitive.
The visited set prevents revisiting any node, so its neighbor list and its vector are retrieved at most once.
Across layers, the search may revisit a logical node because each layer builds a separate, increasingly refined neighborhood structure, and the same node provides different information at each layer. So its data may be reused along the search path. 
For graph access, this causes no address repetition: each layer stores its own graph with independently reshuffled addresses, so expanding the same logical node in different layers always reads unrelated locations.
For vector access, however, standard HNSW implementations keep all vectors in one global table for space and cache efficiency, so re-evaluating a node across layers hits the same physical address. 
To avoid this, we assign each layer its own independently reshuffled vector table of size $N_\ell$. 
Notably, within a single layer, the graph and vector tables are reshuffled independently, so a node’s graph and vector entries occupy unrelated physical addresses, avoiding graph-to-vector transitions by reusing an address. 
With this per-layer duplication and independent reshuffling, both graph and vector access become fully non-repetitive.

\begin{figure*}[htbp]
    \centering
    \papergraphics[width=\linewidth,trim=0 2.83bp 0 12.82bp,clip]{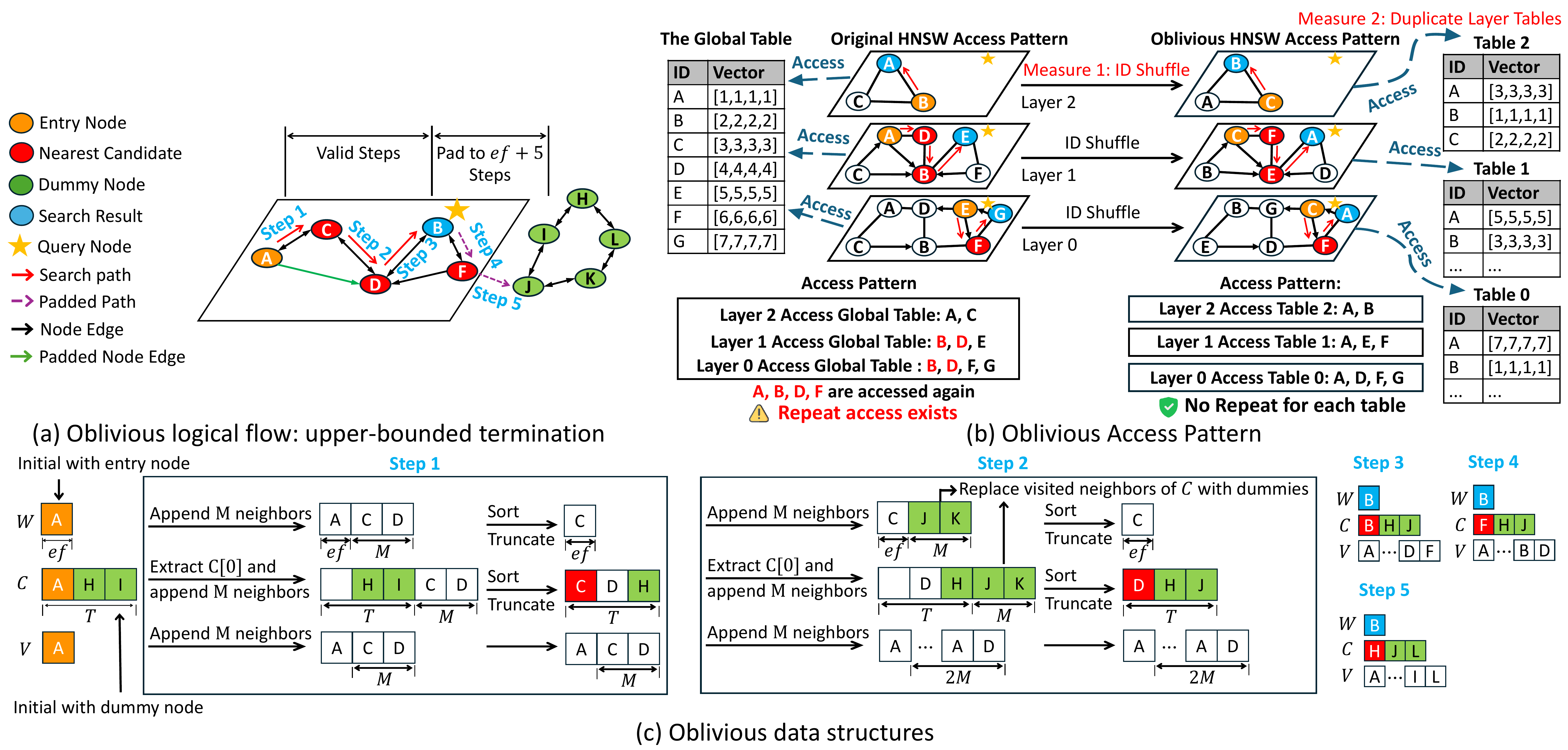}
    \caption{\textbf{Overview of Oblivious HNSW.}
(a)~Each layer runs exactly $\mathit{ef}+\Delta$ steps for data-independent
termination.
(b)~Per-layer replication and reshuffling hide cross-layer access patterns.
(c)~Oblivious data structures maintain the HNSW search state.}
    \label{fig:oblivious_access_pattern}
\end{figure*}

\subsection{Put It All Together}
\label{sec:put_it}

We now integrate the oblivious control flow, data structures, and access patterns into a complete single query HNSW protocol under MPC. Here, we first describe how the per–layer data is represented and secret–shared.
At each layer $\ell$, the data owner provides two tables: a graph table
$\mathcal{G}_\ell$ containing neighbor lists as the HNSW index and a vector table
$\mathcal{D}_\ell$ containing the node vectors.  
Both tables are padded with $TM{+}T{+}\mathit{ef}$ dummies, where $TM$ is for visited check, while $T+\mathit{ef}$ is for initialization of $C$ and $W$.
After padding, MPC parties secret–share and jointly reshuffle the tables offline.
Each graph table entry stores $M$ neighbors' vector-table addresses, $\mathcal{G}_\ell[i] = (\share{n_{i,0}},\ldots,\share{n_{i,M-1}})$, where $\mathcal{D}_\ell[n_{i,j}]$ would return the $j$-th neighbor's vector and auxiliary information. Each vector table entry is $\mathcal{D}_\ell[j] = (\share{\vec{v_j}},\ \share{a_j},\ \share{b_j},\ \share{m_j})$, where $\vec{v}_j$ is the node's vector; $a_j$ is the graph table address of the nodes; $b_j$ is the address of the same vector in $\mathcal{D}_{\ell-1}$ (used during layer transition); and $m_j\in\{0,1\}$ is the dummy entry flag. We collect all dummy vector addresses in a pool $\mathrm{DUM}_\ell = (\share{d_0}, \ldots, \share{d_{MT+T+\mathit{ef}-1}})$.

During the search, MPC parties jointly maintain three fixed-size arrays at each layer: $V = \{\share{a_i}\}_{i=0}^{MT}$, $C = ((\share{c_i},\share{d_i}))_{i=0}^{T-1}$, and $W = ((\share{w_i},\share{b_i},\share{\delta_i}))_{i=0}^{\mathit{ef}-1}$.
Here  
$V$ stores the vector-table addresses $a_i$ of visited nodes;  
$C$ stores the graph-table addresses $c_i$ and distances $d_i$ of candidate nodes;  
$W$ stores the vector-table address $w_i$, the lower-layer vector address $b_i$, and the distance $\delta_i$ of the current top-$\mathit{ef}$ results nodes.

With the oblivious data structures in place, Algorithm~\ref{alg:secure-search-layer} specifies the per-layer search, which runs a fixed number of iterations and updates $V$, $C$, and $W$ obliviously. The full single-query protocol (Algorithm~\ref{alg:mpc-hnsw-search}) applies this layer search from the top layer to the base. At the base layer, $W_0$ stores only vector-table addresses for efficient oblivious sorting; the actual vectors are fetched from an auxiliary reshuffled table $\mathcal{D}_{-1}$. Since the base layer may place dummy addresses among the top-$k$ when $k$ is large, $\mathcal{D}_{-1}$ is padded with $MT{+}T{+}\mathit{ef}$ dummy entries containing application-safe vectors so that dummy outputs remain indistinguishable from real ones.
\begin{algorithm}[!t]
\caption{\textsc{ObliviousSearchLayer}: Oblivious layer search}
\label{alg:secure-search-layer}
\small
\begin{algorithmic}[1]
\Require Query $\vec{q}$, entry point $\mathit{ep}$, search width $\mathit{ef}$, slack $\overline{\Delta}^{(\ell)}$, degree bound $M$, layer $\ell$ (per-layer tables $G_\ell, D_\ell, \mathrm{DUM}_\ell$ accessed implicitly)
\Ensure Set $W$ of $\mathit{ef}$ closest neighbors to $\vec{q}$ in layer $\ell$

\State $T \gets \mathit{ef} + \overline{\Delta}^{(\ell)}$
\State initialize visited array $V[0:T \cdot M] \gets \sigma_{\mathrm{addr}}$ \Comment{dummy graph address}
\State initialize candidate array $C[0:T+M-1] \gets (\sigma_{\mathrm{addr}}, d_{\max})$
\State initialize result array $W[0:\mathit{ef}+M-1] \gets (\sigma_{\mathrm{addr}}, \sigma_{\mathrm{addr}}, d_{\max})$

\State $(\vec{v}_{\mathit{ep}}, a_{\mathit{ep}}, b_{\mathit{ep}}, m_{\mathit{ep}}) \gets D_\ell[\mathit{ep}]$
\State $d_{\mathit{ep}} \gets \mathit{dist}(\vec{v}_{\mathit{ep}}, \vec{q})$
\State $V[0] \gets a_{\mathit{ep}}$
\State $C[0] \gets (a_{\mathit{ep}}, d_{\mathit{ep}})$
\State $W[0] \gets (\mathit{ep}, b_{\mathit{ep}}, d_{\mathit{ep}})$

\For{$t = 0$ \textbf{to} $T-1$}
    \State $(c, d_c) \gets C[0]$ \Comment{current best candidate}
    \State shift $C[1:T-1]$ left by one position
    \State $C[T-1] \gets (\sigma_{\mathrm{addr}}, d_{\max})$
    \State $(n_0, \ldots, n_{M-1}) \gets G_\ell[c]$ \Comment{neighbors of $c$ in layer $\ell$}

    \For{$i = 0$ \textbf{to} $M-1$}
        \State $u \gets n_i$
        \State $\mathit{visited} \gets \textsc{LinearScan}(V, u)$
        \State $\textit{dummy} \gets \mathrm{DUM}_\ell[T + \mathit{ef} + t \cdot M + i]$
        \State $u' \gets \textsc{Select}(\mathit{visited}, \textit{dummy}, u)$
        \State $(\vec{v}_u, a_u, b_u, m_u) \gets D_\ell[u']$
        \State $\mathit{mask} \gets \mathit{visited} \lor m_u$
        \State $d_u \gets \textsc{Select}(\mathit{mask}, d_{\max}, \mathit{dist}(\vec{v}_u, \vec{q}))$
        \State $V[t \cdot M + i + 1] \gets a_u$
        \State $C[T + i] \gets (a_u, d_u)$
        \State $W[\mathit{ef} + i] \gets (u', b_u, d_u)$
    \EndFor

    \State $C \gets \textsc{ObliviousSortTruncate}(C, T)$ \Comment{keep best $T$ of $T{+}M$}
    \State $W \gets \textsc{ObliviousSortTruncate}(W, \mathit{ef})$ \Comment{keep best $\mathit{ef}$ of $\mathit{ef}{+}M$}
\EndFor

\State \Return the first $\mathit{ef}$ entries of $W$
\end{algorithmic}
\end{algorithm}

\subsubsection{Security Sketch}
Throughout the protocol, the query, the graph and vector tables, and the search state remain secret-shared. For fixed public parameters, each query runs the same number of iterations and updates the fixed-size arrays $V$, $C$, and $W$ in the same way. Thus, the logic flow, data-structure sizes, and update schedule do not depend on private inputs. The graph and vector tables are padded and randomly reshuffled before the search. Each lookup accesses a distinct physical address. The observed access sequence is therefore a random sequence of distinct addresses. Thus, the protocol securely realizes the ideal functionality in Figure~\ref{func}. See the extended version for the full proof~\cite{hnsecw_extended}.

\subsubsection{Complexity Analysis}
We analyze the online complexity of the single-query protocol. At the base layer, we set $\mathit{ef}=O(k)$ and run $T=O(k)$ iterations. Each iteration performs: $M=O(1)$ distance evaluations in $\mathbb{R}^d$; $M$ oblivious scans over a list $V$ of size $O(k)$; the oblivious sort network evaluation on $C$ and $W$, each of size $O(k)$. Over $T$ iterations, both communication and local computation  are
$O\left(kd+k^2\log^2 k\right)$.
For $O(\log N)$ non-base layers, with $\mathit{ef}=1$, the cost becomes $O(d\log N)$.
Therefore, a single oblivious HNSW query has communication and local computation costs $O(kd+k^2\log^2 k+d\log N)$.

%% file: Chapters/5_Multiple_Query.tex
\section{Multiple Query Protocol}
\label{sec:multi}

In this section, we extend the single-query protocol to support $U$ queries. 
Recall that the single-query protocol searches obliviously over tables pre-shuffled offline. 
A naive approach for $U$ queries is to run the single-query protocol $U$ times, reshuffling the tables after each query, but this incurs prohibitive online costs due to the $(U-1)$ additional reshuffles. 
We therefore aim to answer all $U$ queries using a single shuffled HNSW index with no further online reshuffling. 
To achieve this, we propose using two global cross-query visited structures, one for graph nodes and the other for vectors, so later queries can safely reuse previously accessed information without re-accessing the underlying tables.

\begin{algorithm}[!t]
\caption{\textsc{ObliviousSingleQuery}: HNSW single query}
\label{alg:mpc-hnsw-search}
\small
\begin{algorithmic}[1]
\Require Query $\vec{q}$; top-layer entry $i_{\mathrm{top}}^{\mathcal{D}}$; tables $\{\mathcal{G}_\ell, \mathcal{D}_\ell\}_{\ell=0}^{L-1}$ and $\mathcal{D}_{-1}$; params $(\mathit{ef},\overline{\Delta},M)$; result count $k$.
\Ensure
  Top-$k$ nearest neighbor vectors of $\vec{q}$.
\State $i_{\mathrm{ep}}^{\mathcal{D}} \gets i_{\mathrm{top}}^{\mathcal{D}}$
\For{$\ell=L-1$ \textbf{to} $1$}
\State ${W_\ell} \gets \textsc{ObliviousSearchLayer}(\vec{q},\, i_{\mathrm{ep}}^{\mathcal{D}},\, \mathit{ef}=1,\,\overline{\Delta}^{(\ell)},\, M,\, \ell)$
  \State $(w^{\mathcal{D}}, b^{\mathcal{D}}, \delta)\gets W_\ell[0]$
  \State $i_{\mathrm{ep}}^{\mathcal{D}} \gets b^{\mathcal{D}}$
\EndFor

\State ${W_0} \gets \textsc{ObliviousSearchLayer}(\vec{q},\, i_{\mathrm{ep}}^{\mathcal{D}},\, \mathit{ef},\,\overline{\Delta}^{(0)},\, M,\, \ell=0)$
\State Reveal $\{b_i^{\mathcal{D}}\}_{i=0}^{k-1}$ in ${W_0}$ \Comment{top-$k$ vector addresses in additional table}
\State Fetch final top-$k$ vectors $\{\vec{v}_i\}_{i=0}^{k-1}=\{\mathcal{D}_{-1}[b_i^\mathcal{D}]\}_{i=0}^{k-1}$
\State Reconstruct $\{\vec{v}_i\}_{i=0}^{k-1}$ to Querier
\State \Return $\{\vec{v}_i\}_{i=0}^{k-1}$
\end{algorithmic}
\end{algorithm}

\subsection{Double Visitor Arrays}
\label{subsec:double_visit}

The challenge in the multiple-query setting is that later queries may need to access nodes and vectors that earlier queries already accessed.
As noted above, the single-query protocol prevents repeated data access with an array $V$ tracking previously accessed nodes.
Moving to the multiple-query setting, one straightforward idea is to simply make $V$ global, so that each node can be accessed by only one query.
However, queries may have overlapping search paths, so simply globalizing $V$ would make multi-query HNSW search incorrect.

To address this issue, we propose two global cross-query caches recording all nodes and vectors accessed so far. Specifically, we maintain a graph-side visited set $V^{\mathcal{G}}_\ell$ and a vector-side visited set $V^{\mathcal{D}}_\ell$, which store the neighbor information of visited nodes and the contents of visited vectors, respectively.
When exploring a node $v$ in graph $\mathcal{G}_\ell$, we first check whether $v \in V^{\mathcal{G}}_\ell$. If so, the neighbor information of $v$ is retrieved from $V^{\mathcal{G}}_\ell$, followed by a dummy graph access. Otherwise, $v$ and all its neighbor information are read from $\mathcal{G}_\ell$ and stored in $V^{\mathcal{G}}_\ell$ for future use.
After obtaining the neighbor information, we access each neighbor's vector to compute its query distance. Similarly, for each vector access, we first check $V^{\mathcal{D}}_\ell$. If the vector is already present, it is retrieved from $V^{\mathcal{D}}_\ell$, followed by a dummy vector access. Otherwise, it is loaded from $\mathcal{D}_\ell$ and appended to $V^{\mathcal{D}}_\ell$.
Moreover, before processing multiple queries, we pre-shuffle both the graphs and the vector databases so that all nodes and vectors, including dummy ones, remain anonymous.
More precisely, for each graph access, we first perform a linear scan of $V^{\mathcal{G}}_\ell$. Then, depending on whether the target node is already in the cache, we access either the real node or a dummy node in the graph. Finally, we access $V^{\mathcal{G}}_\ell$ again to store the newly retrieved real node or a dummy entry. Notably, we generate $U\times T$ dummy nodes so that each graph access can use a different dummy node.
Similarly, for each vector access, we first scan $V^{\mathcal{D}}_\ell$, then access either a real or dummy vector, and finally store it in $V^{\mathcal{D}}_\ell$.

\subsubsection{Security Sketch}
The multi-query protocol extends the single-query protocol with cross-query caches for repeated graph and vector accesses. We explain why this extension remains oblivious. To hide whether a lookup hits a cache, both cases follow the same fixed schedule.
With this design, the control flow is identical for every graph access. Regarding the access pattern, all nodes, including dummy nodes, are accessed at most once, thereby guaranteeing non-repetitive access. In addition, the graph is pre-shuffled to ensure data anonymization.
By the same argument, these vector accesses are also non-repetitive and data-anonymized, and hence oblivious.
Thus, the multi-query protocol securely realizes $U$ sequential invocations of the ideal functionality in Figure~\ref{func}. The full proof is in the extended version~\cite{hnsecw_extended}.

\subsubsection{Complexity Analysis}
For $U$ queries, at the base layer we keep $\mathit{ef}=O(k)$ and run $T=O(k)$ iterations per query. Each iteration performs: $M=O(1)$ distance evaluations in $\mathbb{R}^d$; $M$ oblivious scans over $V^\mathcal{D}$ and $V^\mathcal{G}$ of size $O(Uk)$; and two oblivious sort network evaluations over $C$ and $W$, each of size $O(k)$. Thus, both communication and local computation costs at the base layer are $O(Ukd + Uk^2\log^2 k + U^2k^2)$, and for the $O(\log N)$ non-base layers with $\mathit{ef}=1$ they are $O((Ud+U^2)\log N)$. Therefore, the multiple-query protocol has total communication and local computation costs $O\bigl(Uk(d + k\log^2 k + Uk) + (Ud+U^2)\log N\bigr)$.

%% file: Chapters/6_Batch_Query.tex
\section{Batch Query Protocol in Parallel}
\label{sec:batch}

The preceding multiple-query protocol processes queries sequentially. We further consider the batch setting, where $U$ queries arrive together. We propose a batch protocol that processes the $U$ queries in parallel, reducing the round complexity from $O(U)$ to $O(\log^2 U)$.

The key idea is to share vectors fetched by any query across all $U$ queries, so every query steps forward together by updating its own $C$ and $W$. Each iteration has two stages. First, the protocol obliviously deduplicates the nearest candidate from each query's $C$, replaces duplicates with dummies, and issues $U$ distinct graph accesses. Second, it obliviously deduplicates the $U \cdot M$ returned neighbor IDs and fetches $U \cdot M$ vectors. Each query evaluates its distance to all $U \cdot M$ fetched vectors, regardless of which query requested each vector. The batch therefore performs $U^2 \cdot M$ distance evaluations in parallel. These distances update the query's own candidate set $C$ and result set $W$. The protocol records all fetched vectors in one global visited array $V$ shared by all queries. Each of the two oblivious deduplication steps sorts $O(U)$ elements, requiring an oblivious sort network of depth $O(\log^2 U)$, which determines the per-iteration round cost. The full algorithm and the round-complexity proof appear in the extended version~\cite{hnsecw_extended}.

%% file: Chapters/7_Evaluation.tex
\section{Evaluation}
\label{sec:evaluation}

In this section, we empirically evaluate \sysname\ and aim to answer the following questions:

\begin{itemize}[wide=0pt,font={}]
  \item \textbf{Search quality.}
  What search quality does \sysname\ achieve, and how does it perform when integrated into downstream RAG?
  \item \textbf{Overhead.}
  What computation and communication overhead does \sysname\ incur, and how does it perform against state-of-the-art secure ANN constructions?
\end{itemize}

\subsection{Experimental Setup}
\label{subsec:exp_setup}
\begin{table}[!htbp]
\caption{ANN datasets used in our evaluation~\cite{aumuller2020ann,simhadri2022bigann}.}
\centering
\small
\setlength{\tabcolsep}{2pt}
\begin{tabular}{lrrl|lrrl}
\toprule
\textbf{Dataset} & \textbf{N} & \textbf{d} & \textbf{Metric} & \textbf{Dataset} & \textbf{N} & \textbf{d} & \textbf{Metric} \\
\midrule
SIFT-1M    & $10^{6}$  & $128$ & $\ell_2$  & DEEP-image & $10^{7}$ & $96$  & Cos.     \\
DEEP-1M    & $10^{6}$  & $96$  & $\ell_2$  & DEEP-100M  & $10^{8}$ & $96$  & $\ell_2$ \\
GIST-1M    & $10^{6}$  & $960$ & $\ell_2$  & BIGANN-1B  & $10^{9}$ & $128$ & $\ell_2$ \\
MNIST-60K  & $6{\times}10^{4}$ & $784$ & $\ell_2$ & SPACEV-1B  & $10^{9}$ & $100$ & $\ell_2$ \\
Fashion-60K& $6{\times}10^{4}$ & $784$ & $\ell_2$ & SSNPP-1B   & $10^{9}$ & $256$ & $\ell_2$ \\
\bottomrule
\end{tabular}

\label{tab:datasets}
\end{table}

\noindent \textbf{MPC Backends.}
\sysname\ can be instantiated with any semi-honest MPC protocol tolerating $t$ corruptions among $s$ servers. We evaluate two common configurations:
\begin{itemize}[wide=0pt,font={}]
    \item \textbf{Semi-honest 2PC ($s=2, t=1$).} Our 2PC uses ABY's semi-honest mixed-protocol backend~\cite{demmler2015aby}. Arithmetic sharing handles distance computation. Boolean sharing handles visited checks. Yao's garbled circuits handle comparisons, sorting, and candidate updates. ABY provides secure conversions among these representations.
    \item \textbf{Semi-honest 3PC ($s=3, t=1$).} Our 3PC uses MP-SPDZ's semi-honest replicated secret-sharing protocol over a 64-bit ring~\cite{keller2020mp}.
\end{itemize}

\noindent \textbf{Hardware and Network Environment.}
All experiments run on a single server with an AMD EPYC 9554P CPU (64 physical cores, 3.1~GHz) and 565~GB of RAM, where each MPC party runs as a separate single-threaded process. We emulate two networks: a LAN with 4000~Mbps bandwidth and 1~ms RTT (a standard MPC LAN setting~\cite{li2025panther,chen2020sanns}), and a WAN with 320~Mbps bandwidth and 50~ms RTT, following typical U.S.\ East--West latency per cloudping.co\footnote{\url{https://www.cloudping.co/}. Accessed: 2026-09-23.}.

\noindent \textbf{Datasets.}
We evaluate on $10$ widely-used ANN benchmarks~\cite{aumuller2020ann,simhadri2022bigann} (Table~\ref{tab:datasets}), spanning $N{\in}[6{\times}10^4,10^9]$, $d{\in}[96,960]$, and both $\ell_2$ and cosine metrics. The selected scale range covers the practical RAG deployment regime. Due to space limitations, the main text reports four representative datasets covering both scales (SIFT-1M, GIST-1M, BIGANN-1B, DEEP-100M); results for the remaining six datasets are provided in the extended version~\cite{hnsecw_extended}. All recall experiments use $10^3$ queries per dataset and report the $95$th percentile.

\begin{figure}[!t]
\centering
\papergraphics[width=\linewidth,trim=0 4.21bp 0 2.50bp,clip]{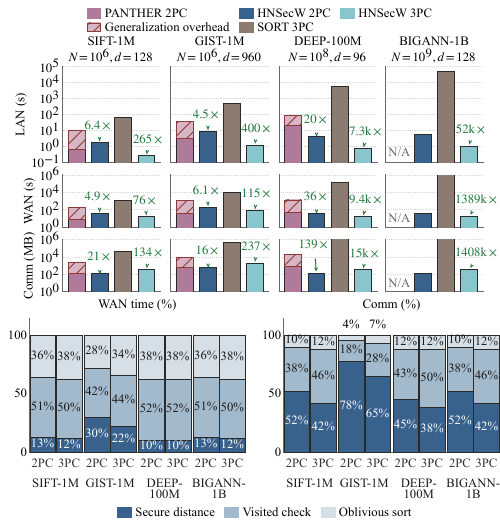}
\caption{\sysname's single-query cost and ablation across the four datasets. In adapted PANTHER, BIGANN-1B exceeds the DORAM~\cite{vadapalli2023gigadoram,vadapalli2023duoram,doerner2017scaling} block-size limit.}
\label{fig:cost-breakdown}
\end{figure}

\noindent \textbf{Comparison Baselines.} As Section~\ref{subsec:mpc} explains, existing secure ANN protocols target different security models and thus provide no direct baseline for our symmetric MPC setting. PANTHER~\cite{li2025panther} is a notable exception: like our protocol, its search algorithm is fully oblivious. However, under its client-server setting, the client accesses the plaintext query and the server stores the plaintext database, enabling lightweight PIR and distance computation between encrypted queries and plaintext vectors by ciphertext–plaintext multiplications. We adapt PANTHER by replacing PIR with DORAM~\cite{vadapalli2023gigadoram} and computing distances by ciphertext–ciphertext multiplications. This adds overhead but preserves its search algorithm. We compare the asymptotic performance of PANTHER's cluster-based search and our graph search under the same security model. Figure~\ref{fig:cost-breakdown} reports PANTHER's original cost and the adaptation overhead separately. Pacmann~\cite{PacmannICLR25} is another related graph-based ANN algorithm for query-private search, but it assumes a public database and graph index, with the client retrieving entries through PIR and searching locally in plaintext. Although PIR can be replaced with DORAM, securely realizing local search would require a complete MPC-based ANN protocol. This is precisely the problem we address. Adapting Pacmann to our setting would be essentially equivalent to implementing our MPC algorithm. For this reason, we do not include it as an experimental baseline. For 3PC, we use replicated radix sort~\cite{asharov2022efficient} as the exact top-$k$ baseline with Recall@$k{=}1$.

\begin{figure}[!t]
\centering
\papergraphics[width=\linewidth,trim=0 4.65bp 0 3.00bp,clip]{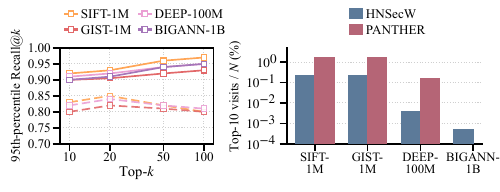}
\caption{Deployed recall of \sysname\ ($M{=}128$,
$\overline{\Delta}^{(\ell)}$ from Section~\ref{subsec:logic_oblivious}). Left: the recall achieved by at least $95\%$ of queries. Right: fraction of database vectors visited at top-$10$. PANTHER is omitted on BIGANN-1B because current DORAMs~\cite{vadapalli2023gigadoram,vadapalli2023duoram,doerner2017scaling} cannot fit its cluster in one block.}
\label{fig:query-performance}
\end{figure}

\begin{figure}[!t]
\centering
\papergraphics[width=\linewidth,trim=0 1.66bp 0 0.00bp,clip]{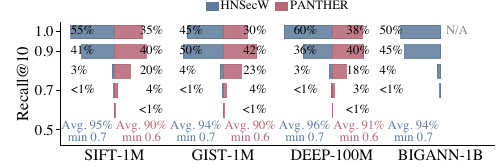}
\caption{Per-query Recall@10 distribution at the deployed setting (1000 queries each).}
\label{fig:recall-distribution}
\end{figure}

\begin{figure}[!t]
\centering
\papergraphics[width=\linewidth,trim=0 2.90bp 0 2.00bp,clip]{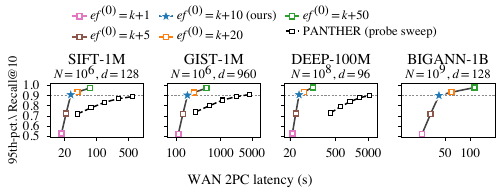}
\caption{$\mathit{ef}^{(0)}$ ablation at $M{=}128$. Each panel sweeps $\mathit{ef}^{(0)} \in \{k{+}1, k{+}5, k{+}10, k{+}20, k{+}50\}$; the star marks our setting $\mathit{ef}^{(0)}{=}k{+}10$ from Section~\ref{subsec:logic_oblivious}. ``probe sweep'' is PANTHER's cluster-probe count $\mathit{n_{probe}} \in \{4, 8, 16, 32, 64\}$.}
\label{fig:ef-ablation-perdata}
\end{figure}

\begin{figure}[!t]
\centering
\papergraphics[width=\linewidth,trim=0 1.62bp 0 3.75bp,clip]{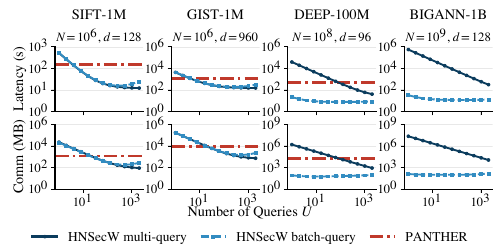}
\caption{Amortized WAN performance of multiple and batch queries with the offline reshuffle.}
\label{fig:mq-batch-combined}
\end{figure}

\subsection{Search Accuracy and RAG Quality}
\label{subsec:eval_accuracy}

\noindent\textbf{Search Recall and Distribution.} At the deployed setting ($M{=}128$, $\mathit{ef}^{(0)}{=}k{+}10$, with $\overline{\Delta}^{(\ell)}$ from Section~\ref{subsec:logic_oblivious}), \sysname\ delivers recall $\ge 0.9$ for at least $95\%$ of queries on every dataset and $k$, with the $95$th-percentile climbing from $\sim 0.9$ at $k{=}10$ to $\sim 0.95$ at $k{=}100$, as shown in Figure~\ref{fig:query-performance}. PANTHER, by contrast, holds Recall@$k\ge 0.8$ for $95\%$ of queries but its $95$th-percentile peaks early near $k{=}20$ and drifts back toward $0.80$ at $k{=}100$: its fixed cluster-probe budget cannot absorb the larger candidate set demanded by larger $k$, while \sysname's fixed-step rule $\mathit{ef}^{(0)}=k+10$ scales with $k$ and pushes the worst-case recall up. \sysname\ also visits $8{\times}$ to $44{\times}$ fewer database vectors than PANTHER as $N$ grows from $10^{6}$ to $10^{8}$ (the range where PANTHER runs), reflecting the $\log N$ vs.\ $\sqrt{N}$ scaling of graph-based vs.\ cluster-based ANN search.
Figure~\ref{fig:recall-distribution} shows the per-query distribution at the deployed setting: $\ge 95\%$ of \sysname's queries land at Recall@$10\ge 0.9$ on every dataset, with $45$--$60\%$ reaching perfect recall, while PANTHER concentrates $\ge 95\%$ of mass at $\ge 0.8$ but only $30$--$38\%$ at perfect recall.
To verify the robustness of the prescribed step parameter, we ablate $\mathit{ef}^{(0)}$ for \sysname\ in Figure~\ref{fig:ef-ablation-perdata}. Recall on every dataset saturates near the deployed $\mathit{ef}^{(0)}{=}k{+}10$ from Section~\ref{subsec:logic_oblivious}: shrinking the offset to $k{+}1$ drops recall sharply, while doubling it to $k{+}20$ or $k{+}50$ buys little extra recall at materially higher WAN latency. PANTHER's recall also rises with the cluster-probe count $\mathit{n_{probe}}$. We further ablate the graph degree $M \in \{16, 32, 64, 128, 256\}$ across all $10$ datasets in the extended version~\cite{hnsecw_extended}; the WAN advantage holds throughout the $M$ sweep.\looseness=-1 %

\begin{table}[!t]
\caption{End-to-end RAG quality on HotpotQA queries. At our setting: $C{=}7$, \sysname\ matches the HNSW baseline.}
\centering
\small
\setlength{\tabcolsep}{2pt}
\begin{tabular}{lcccccccc}
\toprule
        & No retrieval & HNSW & \multicolumn{6}{c}{\sysname} \\
\cmidrule(lr){4-9}
        &      &     & $C{=}1$ & $C{=}3$ & $C{=}5$ & $C{=}7$ & $C{=}9$ & $C{=}11$ \\
\midrule
EM (\%) & 22.4 & 66.2 & 56.8 & 63.5 & 65.2 & \textbf{66.0} & 66.1 & 66.2 \\
F1 (\%) & 31.5 & 81.5 & 70.2 & 78.5 & 80.6 & \textbf{81.3} & 81.4 & 81.5 \\
\bottomrule
\end{tabular}

\label{tab:rag_results}
\end{table}

\noindent\textbf{End-to-end RAG Quality.} We evaluate end-to-end RAG on 200 multi-hop queries sampled from HotpotQA~\cite{yang2018hotpotqa} with retrieval over the 1M-document MS MARCO passage corpus. Queries and documents are embedded with \texttt{intfloat/\allowbreak e5-large-v2}~\cite{wang2022text}, we retrieve the top-10 passages, and \texttt{Qwen2.5-\allowbreak 7B-Instruct}~\cite{bai2025qwen2} generates answers, with Exact Match and token-level F1 as metrics. We sweep $C$ in Table~\ref{tab:rag_results} to characterize RAG quality as $C$ varies. It shows that \sysname\ matches the standard HNSW at $C{=}7$ and larger $C$ gives negligible improvement, so our termination strategy in Section~\ref{subsec:logic_oblivious} achieves similar RAG quality to standard HNSW.

\subsection{Efficiency of \sysname}
\label{subsec:eval_efficiency}

We next compare \sysname's efficiency and scaling for single, multiple, and batch queries.

\subsubsection{Single-Query Scenario}
\label{subsubsec:eff_single}

We first examine single-query cost on the four datasets, shown in Figure~\ref{fig:cost-breakdown}. In 2PC, \sysname\ is $4$–$22\times$ faster than PANTHER on WAN and reduces communication by $7$–$46\times$ (excluding BIGANN-1B, where existing DORAM cannot fit a PANTHER cluster; see Figure~\ref{fig:cost-breakdown} caption). The gap widens against SORT in 3PC, reaching $111{\times}$ to $1.6\mathrm{k}{\times}$ on WAN and $185{\times}$ to $617{\times}$ in communication. The ablation in the bottom row of Figure~\ref{fig:cost-breakdown} decomposes WAN time and Communication into Secure distance, Visited check, and Oblivious sort. WAN is round-dominated, since Visited check and Oblivious sort account for $60\%$ to $90\%$. The Secure distance share rises with $d$, from about $10\%$ at DEEP-100M ($d{=}96$) to about $30\%$ at GIST-1M ($d{=}960$) under 2PC. This share is smaller in 3PC because replicated sharing reduces distance-computation overhead. For Communication, distance bytes dominate at high $d$, reaching about $78\%$ on GIST-1M 2PC, while Visited check and Oblivious sort take larger shares at low $d$.

We further check whether \sysname\ keeps its WAN advantage across the $\mathit{ef}^{(0)}$ ablation in Figure~\ref{fig:ef-ablation-perdata}, which plots WAN 2PC latency against top-$10$ recall. At the same recall, \sysname's 2PC WAN curve sits below PANTHER's $\mathit{n_{probe}}$ sweep on every dataset where PANTHER runs across the full $\mathit{ef}^{(0)} \in \{k{+}1,\ldots,k{+}50\}$ sweep, so the WAN advantage is robust to the choice of $\mathit{ef}^{(0)}$.

\subsubsection{Multiple- and Batch-Query Scenarios}
\label{subsubsec:eff_multi}

We sweep $U$ under WAN and report per-query amortized latency and communication in Figure~\ref{fig:mq-batch-combined}, combining online search with offline reshuffle divided by $U$. At small $U$ the reshuffle dominates; as $U$ grows its share falls linearly while the online share grows slowly, so the amortized total drops toward the online floor. \sysname\ crosses below PANTHER once $U$ amortizes the offline cost. The batch variant starts below the multi-query curve at small $U$ by sharing one parallel reshuffle. The crossover $U$ varies with $N$ and $d$, but all four datasets follow the same trend.

%% file: Chapters/8_Related_Work.tex
\section{Related Work}
\label{sec:related_work}

\textbf{Approximate Nearest Neighbor (ANN).}
ANN enables scalable similarity search in high-dimensional embedding spaces where exact $k$NN is often too expensive~\cite{ann_survey_1, ann_survey_2, ann_survey_3}. Classical methods include tree-based partitioning~\cite{kd_tree}, locality-sensitive hashing~\cite{indyk1998approximate}, and vector quantization~\cite{jegou2010product}. Graph-based indexes, such as HNSW \cite{malkov2018efficient} and NSG~\cite{fu2019fast}, dominate in practice, building proximity graphs for greedy query traversal. Recent systems extend graph-based ANN with improved indexing and retrieval strategies~\cite{jiang2025falcon, wang2024starling, zhu2025hybrid}, while vector databases deploy these methods at scale~\cite{wang_milvus_2021, guo2022manu, chen2024singlestorev}.

\noindent \textbf{Secure ANN.}
Prior work considers secure ANN under several threat models.
When the database is public but the query is private,
Tiptoe~\cite{tiptoe} searches a small candidate subset selected via clustering, trading accuracy for sublinear bandwidth and latency.
Wally~\cite{wally} further improves this pipeline's throughput by batching queries, while Preco~\cite{preco} reduces communication using two non-colluding servers.
Pacmann~\cite{PacmannICLR25} goes beyond clustering-based ANN methods by building a graph-index under query privacy.
When both the database and query are private under a single owner (encrypted outsourcing), Compass~\cite{zhu_compass_2025} directly uses ORAM only to hide the server-side data access pattern, while executing the search logic in plaintext on the client.
Instead, HERS~\cite{engelsma2022hers} scans the encrypted database with FHE at high computational cost.
SANNS~\cite{chen2020sanns} and PANTHER~\cite{li2025panther} use secure two-party computation between a server owning the database and a client owning the query. SANNS proposes a clustering-based secure ANN with optimized distance computation and top-$k$ selection, while PANTHER further improves end-to-end performance.
In contrast, \sysname\ secret-shares the query, database, and index among computation servers and achieves the same logarithmic complexity as the non-private setting.
Notably, both SANNS~\cite{chen2020sanns} and PANTHER~\cite{li2025panther} can be adapted to secret-shared inputs. However, both approaches have $O(\sqrt{N})$ complexity. Moreover, several early works~\cite{songhori2015compacting,shaul2020secure,schoppmann2018private,zuber2021efficient} studied secure $k$NN protocols, but they either target smaller-scale settings or apply only to the specialized secure settings discussed above.

\section{Conclusion}
We present \sysname, an MPC ANN protocol with $O(\log N)$ online time and symmetric privacy for the query and database. It supports single-query, sequential multi-query, and batch-query settings.
Future work includes more efficient secure index construction and dynamic graph updates. Currently, the database owner builds the index locally before queries arrive. For secure offloading, MPC servers can insert the $N$ vectors into the graph one by one and reuse our oblivious search to find neighbors. This requires $N$ graph searches, $O(N\log N)$ time, and $O(NM)$ space for degree bound $M$~\cite{malkov2018efficient}.

%% file: Chapters/Acknowledgments.tex
\begingroup
\renewcommand{\acksname}{Acknowledgments}
\begin{acks}
This research is supported by the Singapore Ministry of Education Tier 1 Grant (RG19/25) and the NTU–NAP Startup Grant (024584-00001). \textcolor{black}{We also thank the anonymous reviewers who have made valuable suggestions on improving the presentation of the paper.}
\end{acks}
\endgroup